\documentclass{article}
\usepackage[utf8]{inputenc}

\usepackage{hyperref}
\usepackage{amsmath,amsfonts,amssymb}
\usepackage{amsthm}
\usepackage{color}
\usepackage{xcolor}
\usepackage{graphicx}
\usepackage{tikz-cd}
\usepackage{gmp}
\usepackage{braket}
\usepackage{bbold}
\usepackage{cite}
\usetikzlibrary {decorations.pathmorphing,shadows,fadings}

\usetikzlibrary{knots}

\def\ds{\displaystyle}
\def\bea{\begin{array}{c}}
\def\ea{\end{array}}
\def\be{\begin{equation}\bea\ds}
\def\ee{\ea\end{equation}}
\def\bee{\begin{equation}\begin{array}{rcl}\ds}
\def\eee{\end{array}\end{equation}}

\def\Hc{{\mathcal{H}}}

\def\Tr{{\rm Tr}\,}

\title{Error-Correcting Codes in TQFT on Multispheres}
\author{Rafael Chaves$^{a}$, Dmitry Melnikov$^{a}$, Marcos Neves$^{b}$, Luigy Pinto$^{b}$ and Davide Poderini$^{a}$}
\date{}

\begin{document}

\maketitle

\begin{center}
\textit{\small $^{a}$International Institute of Physics, Federal University of 
Rio Grande do Norte, \\ Campus Universit\'ario, Lagoa Nova, Natal-RN  
59078-970, Brazil}

\vspace{0.6cm}
\textit{\small $^{b}$Department of Theoretical and Experimental Physics, Federal University of Rio Grande do Norte, \\ Campus Universit\'ario, Lagoa Nova, Natal-RN  
59078-970, Brazil}

\vspace{2cm}

\end{center}

\vspace{-2cm}

\begin{abstract}
Topological quantum field theories (TQFT) encode quantum correlations in topological features of spaces. In this work, we leverage this feature to explore how information encoded in TQFTs can be stored and retrieved in the presence of local decoherence affecting its physical carriers. TQFT states' inherent nonlocality, redundancy, and entanglement position them as natural error-correcting codes. We demonstrate that information recovery protocols can be derived from the principle that protected information must be uniformly distributed across the system and from interpreting correlations in terms of space connectivity. Specifically, we employ a topological framework to devise erasure error correction protocols, showing that information can be successfully recovered even when parts of the system are corrupted.
\end{abstract}

\bigskip

\section{Introduction}

Topological quantum field theories (TQFT)~\cite{Witten:1988ze,Atiyah:1989vu} construct representations of quantum states and operators in terms of topological spaces. Such a representation encodes various properties of quantum mechanics in the topological features of the spaces. This, in turn, allows for different forms of diagrammatic and visual interpretation of quantum processes. The approach is related to the categorical quantum mechanics~\cite{Abramsky:2004doh,Duncan:2009ocf,Abramsky:2009cat,Coecke:2017dti} and provides an explicit example of ``quantum picturalism''~\cite{Coecke:2010qua}. A great advantage of the TQFT realization in comparison to other examples of picturalism is that it allows to build perhaps the simplest self-contained description of quantum mechanics. 

In the present work, we will elaborate on the above statements and give further support to them by exploring the way the information is stored in TQFT states. We will focus on a specific setup discussed in~\cite{Melnikov:2018zfn,Melnikov:2022qyt} derived from the general axioms of TQFT~\cite{Atiyah:1989vu} and the explicit example provided by Chern-Simons theory~\cite{Witten:1988hf}. Topological constraints on the physical realizations of TQFT, such as anyons~\cite{Halperin:1984fn,Arovas:1984qr,Wen:1990iu} or Majorana fermions~\cite{Fu:2007moz,Nayak:2008zza}, require the information to be stored nonlocally and redundantly, allowing for its recovery if a part of the physical storage is lost or gets corrupted. Specifically, we will discuss how an input logical qubit, encoded in a set of physical qubits, can be recovered if access to one of the physical qubits is lost.

Our approach begins with a standard strategy commonly employed in error correction and secret sharing protocols~\cite{Cleve:1999qg}, but here we adapt it to topologically encoded qubits. The process involves encoding a logical qubit across multiple physical qubits, ensuring that the information (i.e., correlations) is uniformly distributed among them. This uniform distribution allows for the recovery of the logical qubit's information, even after erasing one of the physical qubits, by performing operations on the remaining qubits. Depending on the specific protocol, recovery will either be guaranteed or subject to a known probability bound.

The realization of qubits within a  TQFT framework enables the construction of protocols through intuitive manipulation of the topology associated with quantum states -- essentially, by drawing diagrams. In these diagrams, quantum correlations are depicted by lines connecting different parts of the system, with quantum operations corresponding to topological manipulations of these lines, such as braiding. Although this approach bears similarities to the braiding of anyons in the quantum Hall effect~\cite{Kitaev:1997wr}, it can be seen as more general. The lines in our diagrams represent quantum correlations that exist even when an anyon interpretation is not applicable. We will first introduce the topological algorithm and then provide an explicit realization on the computational basis. Given the specific construction of TQFT Hilbert spaces, we will consider two versions of the encoding and recovery algorithms.

The first one involves working with Hilbert spaces constructed as tensor products of individual qubits. In this otherwise natural scenario, we encounter difficulties in recovering information with certainty. These difficulties are related to the intrinsic properties of the TQFTs. For this case, we assume the availability of multiple certified copies of the encoded qubit and derive bounds on the probability of a successful recovery. We find that probability is around 30\%, almost independent from the encoded state and from the choice of the TQFT. With this somewhat unimpressive success rate, we believe that more interesting in this example is the technology behind the construction of the protocol.

The limitations of the first protocol can be addressed by embedding qubits in a larger Hilbert space without a tensor product structure, which is a more natural setup for physical implementations such as anyons~\cite{Nayak:2008zza,Kitaev:1997wr}. In this alternative approach, the second version of the protocol allows for the information to be recovered with certainty, at least in principle. In one specific scenario, we construct an explicit unitary transformation that recovers the information stored in the spins encoding the topological qubit. In a more general situation, the recovery process may further depend on the specifics of the physical realization. 

TQFTs are believed to provide an effective description of quantum states in phases of matter classified as topological. The primary interest in these states stems from their robustness against decoherence, a direct consequence of their topological properties. This robustness has led to a substantial body of literature on topological quantum computation and its many applications (see reviews~\cite{Nayak:2008zza,Aasen:2016qpo,Field:2018qtg} for details and references). This work begins with a complementary perspective on topological quantum computation by exploring the general properties of TQFTs and attempting to formulate fundamental principles for manipulating quantum states using natural topological tools~\cite{Melnikov:2017bjb,Melnikov:2018zfn,Melnikov:2022qyt,Melnikov:2023nzn}, rather than the standard approaches involving bases, gates, and quantum circuits. A similar discussion regarding error-correcting codes was recently presented in~\cite{Fields:2022zvj,Fields:2023uqw,Fields:2024ctw}, where it was shown that TQFTs emerge from Quantum Reference Frames and thus provide representations of multiparty communication protocols inducing quantum error correction codes. One important aspect discussed there is the natural classical redundancy emerging from spacetime in the context of TQFTs. Our work explores similar ideas on a more technical level, focusing on a specific problem of erasure correction and using an explicit presentation of TQFT states.

The discussion of error-correcting codes within the TQFT framework fits naturally in the broader discussion on holography — the duality between conformal field theories and quantum gravity in anti-de Sitter spaces~\cite{Maldacena:1997re,Aharony:1999ti,Harlow:2014yka} — as an error-correcting code~\cite{Almheiri:2014lwa,Verlinde:2012cy}. In holography, higher-dimensional quantum gravity states are encoded on the boundary of the space in a manner that protects the information against local erasures. The protocols developed in this paper share a similar spirit and can be compared with tensor networks, which have been constructed as models of holography~\cite{Pastawski:2015qua,Harlow:2014yka,Yoshida:2017non,Akers:2022qdl}. Formal language for the discussion of error-correcting codes in this context involves von Neumann algebras~\cite{Harlow:2016vwg,Kang:2019dfi,Faulkner:2020iou,Faulkner:2020hzi,Gesteau:2021jzp}, also bearing connections with TQFTs.

This paper is organized as follows. In section~\ref{sec:tqubits} we introduce topological qubits and their Hilbert spaces. In section~\ref{sec:topcorrection} we present two topological protocols for encoding the logical qubit in four physical ones and recovery of the logical qubit after a loss of one physical qubit. 

The first protocol is sketched in section~\ref{sec:errcorr1}. In this protocol, the space of physical qubits is simply a direct product of spaces of individual qubits. It turns out that in such a setting there are no natural topological operations, such as braiding, that could be used for information recovery. Naive braiding includes projections and therefore is a probabilistic operation. However, stochastically, the information can be recovered in a very elegant way by a pair of measurements. We calculate the probabilities of a successful recovery for this most straightforward protocol.

The second protocol, considered in section~\ref{sec:errcorr2} embeds physical qubits into a larger space, preserving the unitarity of braiding. Braiding can be used to concentrate the distributed information locally, although the presence of correlations with the corrupted region may potentially prevent the full recovery. To show that these correlations pose no problem, and the full information is available locally, we turn to an explicit description of physical qubits in terms of (pseudo) spins, which is discussed in section~\ref{sec:errcorr3}. In section~\ref{sec:punitary} we review a specific pseudounitary representation of TQFT quantum mechanics embedded in the tensor product of two-dimensional Hilbert spaces (spins). We use this representation in section~\ref{sec:encoding} to illustrate how a single topological qubit is encoded in four physical spins. In particular, we show that losing one physical spin permits distilling the encoded logical qubit in any of the three remaining spins. 

We discuss our results and give concluding remarks in section~\ref{sec:conclusions}.

\section{Topological qubits}
\label{sec:tqubits}

In this work, generic quantum states will be represented by topological 3-manifolds with 2-sphere boundaries.\footnote{See the general TQFT axioms explaining how topological spaces encode quantum states in Refs.~\cite{Atiyah:1989vu,Kauffman:2013bh}.} Each sphere will have a number of marked points (punctures), that can be identified with \emph{anyons}, connected between each other by 1-dimensional defects, called Wilson lines~\cite{Witten:1988hf}, using the terminology of the Chern-Simons TQFTs. A Hilbert space factor will be associated with each sphere -- a qudit, whose dimension will depend on the number of anyons. Here is an example of one-qudit and two-qudit states (to be detailed later):
\be
\label{topqubits}
\begin{array}{c}
     \begin{tikzpicture}
         \draw[line width=1.5] (-0.2,-0.4) .. controls +(0.15,0.35) .. (0.3,-0.25);
          \draw[line width=1.5] (-0.4,0.2) .. controls +(0.45,-0.25) .. (0.1,0.35);
         \fill[olive,opacity=0.7] (0,0) circle (0.75);
         \draw[gray] (0.75,0) arc (0:-250:0.75);
         \fill[white,opacity=0.7,rounded corners=1] (0.4,0.225) -- (0.475,0.25) -- (0.425,0.425) -- (0.325,0.4) -- (0.35,0.225) -- (0.35,0.225); 
         \fill (-0.2,-0.4) circle (0.075);
         \fill (0.3,-0.25) circle (0.075);
         \fill (-0.4,0.2) circle (0.075);
         \fill (0.1,0.35) circle (0.075);
     \end{tikzpicture} 
\end{array}
\ \equiv \  \begin{array}{c}
     \begin{tikzpicture}
        \draw[draw=white,double=black,line width=1,rounded corners=2] (-0.45,0) -- (-0.45,-0.7) -- (0.45,-0.7) -- (0.45,0);
        \draw[draw=white,double=black,line width=1,rounded corners=2] (0.15,0) -- (0.15,-0.5) -- (-0.15,-0.5) -- (-0.15,0);
         \fill[olive,opacity=0.8,rounded corners=3] (0,-.2) -- (0.75,-.2) -- (0.75,0.2) -- (-0.75,0.2) -- (-0.75,-.2) -- (0,-.2);
         \foreach \x in {-0.45,-0.15,...,0.75}
           \fill[black] (\x,0) circle (0.075);
     \end{tikzpicture} 
\end{array} \qquad\qquad
\begin{array}{c}
     \begin{tikzpicture}
         \fill[gray,opacity=0.8] (-0.1,0) circle (0.3);
         \draw (0.2,0) arc (0:-250:0.3);
         \fill (-0.15,-0.1) circle (0.05);
         \fill (-0.2,0.05) circle (0.05);
         \fill (0.,-0.05) circle (0.05);
         \fill (-0.05,0.1) circle (0.05);
         \draw[line width=1.5] (-0.25,-0.5) .. controls +(-0.2,0.25) .. (-0.15,-0.1);
         \draw[line width=1.5] (-0.4,0.3) .. controls +(-0.2,-0.25) .. (-0.2,0.05);
         \draw[line width=1.5] (0.3,-0.25) .. controls +(0.1,0.25) .. (0.0,-0.05);
         \draw[line width=1.5] (0.1,0.35) .. controls +(-0.2,0) .. (-0.05,0.1);
         \fill[olive,opacity=0.7] (0,0) circle (0.75);
         \draw[gray] (0.75,0) arc (0:-250:0.75);
         \fill[white,opacity=0.7,rounded corners=1] (0.4,0.225) -- (0.475,0.25) -- (0.425,0.425) -- (0.325,0.4) -- (0.35,0.225) -- (0.35,0.225); 
         \fill (-0.25,-0.5) circle (0.075);
         \fill (0.3,-0.25) circle (0.075);
         \fill (-0.4,0.3) circle (0.075);
         \fill (0.1,0.35) circle (0.075);
     \end{tikzpicture} 
\end{array}
\ \equiv \  \begin{array}{c}
     \begin{tikzpicture}
     \newcommand{\y}{2}
        \draw[draw=white,double=black,line width=1,rounded corners=2] (-0.45,0) -- (-0.45,-0.7) -- (0.45+\y,-0.7) -- (0.45+\y,0);
        \draw[draw=white,double=black,line width=1,rounded corners=2] (-0.15,0) -- (-0.15,-0.6) -- (0.15+\y,-0.6) -- (0.15+\y,0);
        \draw[draw=white,double=black,line width=1,rounded corners=2] (0.15,0) -- (0.15,-0.5) -- (-0.15+\y,-0.5) -- (-0.15+\y,0);
         \draw[draw=white,double=black,line width=1,rounded corners=2] (0.45,0) -- (0.45,-0.4) -- (-0.45+\y,-0.4) -- (-0.45+\y,0);
         \fill[olive,opacity=0.8,rounded corners=3] (0,-.2) -- (0.75,-.2) -- (0.75,0.2) -- (-0.75,0.2) -- (-0.75,-.2) -- (0,-.2);
         \foreach \x in {-0.45,-0.15,...,0.75}
           \fill[black] (\x,0) circle (0.075);
          \fill[olive,opacity=0.8,rounded corners=3] (0+\y,-.2) -- (0.75+\y,-.2) -- (0.75+\y,0.2) -- (-0.75+\y,0.2) -- (-0.75+\y,-.2) -- (0+\y,-.2); 
          \foreach \x in {-0.45,-0.15,...,0.75}
           \fill[black] (\x+\y,0) circle (0.075);
     \end{tikzpicture} 
\end{array}
\ee
In the second example the lines connect a pair of concentric spheres. To simplify the drawing, we pass to the heuristic depiction of spaces, which only features their boundaries and the Wilson lines piercing the 3-dimensional spaces.

This kind of correspondence between states and diagrams is common for different three-dimensional TQFT's. Here we will specify the details in the case of Chern-Simons theory with gauge group $SU(2)$ and coupling constant $k$~\cite{Witten:1988hf,Melnikov:2022qyt}. The anyons will correspond to external nondynamical particles in the fundamental representation of $SU(2)$. The dimension of the qudit will be determined by the number of ways that the anyons can form a spin-singlet state. For four fundamental anyons, there are two such ways, so the diagrams in equation~(\ref{topqubits}) actually show one and two-qubit states. More precisely, the dimension also depends on the value of $k$. For integer $k$, the dimension can be lower than the number of spin singlet combinations. This happens because some states in the Hilbert space become null, or, equivalently some diagrams become linear combinations of other diagrams. If $k$ is sufficiently large, the dimension of the Hilbert space of a sphere with $2n$ punctures $\Hc_{2n}$ is equal to the $n^{\rm th}$ Catalan number\footnote{The problem of counting singlets can be mapped to the problem of counting the dimension of the Temperley-Lieb algebra $TL_n$~\cite{kassel2008braid} or to a number of equivalent combinatorial problems.}
\be
\label{Catalan}
\dim \Hc_{2n} \ = \ C_n \ = \ \frac{(2n)!}{(n+1)!n!}\,, \qquad k>n-1\,.
\ee

We will generally assume that $k$ is sufficiently large and the qudit dimension is given by $C_n$. For spheres with four anyons, that is, qubits with $k\neq 1$, any pair of topologically inequivalent diagrams provides a basis. It is convenient to choose the following pair:
\be
\label{basis}
|\hat0\rangle\ =\ 
\begin{array}{c}
     \begin{tikzpicture}
        \draw[draw=white,double=black,line width=1,rounded corners=2] (-0.45,0) -- (-0.45,-0.6) -- (-0.15,-0.6) -- (-0.15,0);
        \draw[draw=white,double=black,line width=1,rounded corners=2] (0.45,0) -- (0.45,-0.6) -- (0.15,-0.6) -- (0.15,0);
         \fill[olive,opacity=0.8,rounded corners=3] (0,-.2) -- (0.75,-.2) -- (0.75,0.2) -- (-0.75,0.2) -- (-0.75,-.2) -- (0,-.2);
         \foreach \x in {-0.45,-0.15,...,0.75}
           \fill[black] (\x,0) circle (0.075);
     \end{tikzpicture}
\end{array}  \,, \qquad |\hat 1\rangle \ = \ \begin{array}{c}
     \begin{tikzpicture}
        \draw[draw=white,double=black,line width=1,rounded corners=2] (-0.45,0) -- (-0.45,-0.7) -- (0.45,-0.7) -- (0.45,0);
        \draw[draw=white,double=black,line width=1,rounded corners=2] (0.15,0) -- (0.15,-0.5) -- (-0.15,-0.5) -- (-0.15,0);
         \fill[olive,opacity=0.8,rounded corners=3] (0,-.2) -- (0.75,-.2) -- (0.75,0.2) -- (-0.75,0.2) -- (-0.75,-.2) -- (0,-.2);
         \foreach \x in {-0.45,-0.15,...,0.75}
           \fill[black] (\x,0) circle (0.075);
     \end{tikzpicture} 
\end{array} .
\ee
This basis is not orthonormal. To construct an orthonormal basis one applies the Gram-Schmidt procedure, constructing linear combinations of diagrams. For this, one needs to compute overlaps of states $|\hat{0}\rangle$ and $|\hat{1}\rangle$.

The overlaps, or inner products, are obtained by gluing the qubit spaces along their boundary. The result of gluing two 3-balls along $S^2$ is a 3-sphere $S^3$. In this gluing, the Wilson lines close, forming a knot or a link inside $S^3$. Then the inner product is given by the Jones polynomial of this knot or link. The latter can be computed using a set of simple rules (e.g.~\cite{Kauffman:2013bh}). First, any disconnected trivial (unknotted) loop can be replaced by a numerical factor $d$. This rule allows to compute all the overlaps of the basis vectors:
\be
\label{overlaps}
\langle \hat{0}|\hat{0}\rangle \ = \  
\begin{array}{c}
     \begin{tikzpicture}
        \draw[draw=white,double=black,line width=1,rounded corners=2] (0,-0.45) -- (-0.6,-0.45) -- (-0.6,-0.15) -- (0,-0.15);
        \draw[draw=white,double=black,line width=1,rounded corners=2] (0,0.45) -- (-0.6,0.45) -- (-0.6,0.15) -- (0,0.15);
         \fill[olive,opacity=0.8,rounded corners=3] (-.2,0) -- (-.2,0.75) -- (0.2,0.75) -- (0.2,-0.75) -- (-.2,-0.75) -- (-.2,0);
         \foreach \x in {-0.45,-0.15,...,0.75}
           \fill[black] (0,\x) circle (0.075);
         \draw[draw=white,double=black,line width=1,rounded corners=2] (0,-0.45) -- (0.6,-0.45) -- (0.6,-0.15) -- (0,-0.15);
        \draw[draw=white,double=black,line width=1,rounded corners=2] (0,0.45) -- (0.6,0.45) -- (0.6,0.15) -- (0,0.15);
     \end{tikzpicture} 
\end{array} \ = \ d^2\,, \qquad
\langle \hat{0}|\hat{1}\rangle \ = \  
\begin{array}{c}
     \begin{tikzpicture}
        \draw[draw=white,double=black,line width=1,rounded corners=2] (0,-0.45) -- (-0.6,-0.45) -- (-0.6,-0.15) -- (0,-0.15);
        \draw[draw=white,double=black,line width=1,rounded corners=2] (0,0.45) -- (-0.6,0.45) -- (-0.6,0.15) -- (0,0.15);
         \fill[olive,opacity=0.8,rounded corners=3] (-.2,0) -- (-.2,0.75) -- (0.2,0.75) -- (0.2,-0.75) -- (-.2,-0.75) -- (-.2,0);
         \foreach \x in {-0.45,-0.15,...,0.75}
           \fill[black] (0,\x) circle (0.075);
        \draw[draw=white,double=black,line width=1,rounded corners=2] (0,-0.45) -- (0.7,-0.45) -- (0.7,0.45) -- (0,0.45);
        \draw[draw=white,double=black,line width=1,rounded corners=2] (0,-0.15) -- (0.5,-0.15) -- (0.5,0.15) -- (0,0.15);
     \end{tikzpicture} 
\end{array} \ = \ d\,, \qquad
\langle \hat{1}|\hat{1}\rangle \ = \  
\begin{array}{c}
     \begin{tikzpicture}
        \draw[draw=white,double=black,line width=1,rounded corners=2] (0,-0.45) -- (-0.7,-0.45) -- (-0.7,0.45) -- (0,0.45);
        \draw[draw=white,double=black,line width=1,rounded corners=2] (0,-0.15) -- (-0.5,-0.15) -- (-0.5,0.15) -- (0,0.15);
         \fill[olive,opacity=0.8,rounded corners=3] (-.2,0) -- (-.2,0.75) -- (0.2,0.75) -- (0.2,-0.75) -- (-.2,-0.75) -- (-.2,0);
         \foreach \x in {-0.45,-0.15,...,0.75}
           \fill[black] (0,\x) circle (0.075);
        \draw[draw=white,double=black,line width=1,rounded corners=2] (0,-0.45) -- (0.7,-0.45) -- (0.7,0.45) -- (0,0.45);
        \draw[draw=white,double=black,line width=1,rounded corners=2] (0,-0.15) -- (0.5,-0.15) -- (0.5,0.15) -- (0,0.15);
     \end{tikzpicture} 
\end{array} \ = \ d^2\,.
\ee
Similarly, the Jones polynomial of a multicomponent link which has a disconnected unknotted circle besides a general link or knot, is equal to the same factor $d$ times the polynomial of that link or knot. Computing the Gram matrix shows that states in the basis~(\ref{basis}) are linearly independent, except for some special values of $d$. Since $d$ is a function of $k$, as specified below, one can see why the actual dimension of the Hilbert space might be smaller than the number~(\ref{Catalan}) given by the combinatorial counting of diagrams.

Finally, to compute the Jones polynomial of a generic link, one can use the skein relations~\cite{Kauffman:1987sta}, which express linear relations between diagrams:
\be
\label{skein}
\begin{array}{c}
     \begin{tikzpicture}
         \draw[line width=1.5,rounded corners=3] (-0.7,-0.3) -- (-0.3,-0.3) -- (0.3,0.3) -- (0.7,0.3);
         \draw[line width=1.5,rounded corners=3] (-0.7,0.3) -- (-0.3,0.3) -- (-0.1,0.1);
         \draw[line width=1.5,rounded corners=3] (0.1,-0.1) -- (0.3,-0.3) -- (0.7,-0.3);
     \end{tikzpicture} 
\end{array} \ = \ A\begin{array}{c}
     \begin{tikzpicture}
         \draw[line width=1.5,rounded corners=3] (-0.7,-0.3) -- (-0.2,-0.3) -- (-0.2,0.3) -- (-0.7,0.3);
         \draw[line width=1.5,rounded corners=3] (0.7,0.3) -- (0.2,0.3) -- (0.2,-0.3) -- (0.7,-0.3);
     \end{tikzpicture} 
\end{array} + A^{-1}\begin{array}{c}
     \begin{tikzpicture}
         \draw[line width=1.5,rounded corners=3] (-0.7,-0.3) -- (0.7,-0.3);
         \draw[line width=1.5,rounded corners=3] (-0.7,0.3) -- (0.7,0.3);
     \end{tikzpicture} 
\end{array}
\,.
\ee
This relation should be understood as a linear relation between three objects (states or operators) that differ only in a specific neighborhood of given two Wilson lines. In other words, any crossing in the diagram can be replaced by a linear combination of two diagrams with lines wired in a way to avoid the crossing. The coefficients in the linear relation are connected to the above parameters as follows: 
\be
\label{params}
A\ =\ e^{i\theta}\ = \ e^{\frac{\pi i}{2(k+2)}} \,,\qquad \text{and}\qquad d \ = \ - A^2 - A^{-2} \ = \ -2\cos\frac{\pi}{k+2}\,.
\ee
Skein relations reduce an arbitrary link diagram to a linear combination of diagrams containing only unlinked unknotted loops. These rules give the unnormalized Kauffman's version of the Jones polynomial (the bracket polynomial). Note that the rules are slightly different if the knot is drawn in a space different from $S^3$, which can easily happen for overlaps of multipartite states. We will not need those rules in the present work. Further details can be found in~\cite{Witten:1988hf,Kauffman:2013bh}.

We close this brief review of the Chern-Simons states with the expression for the orthonormal basis to be used in this work:
\be
\label{ONbasis}
|0\rangle \ = \ \frac{1}{d}\,|\hat{0}\rangle\,, \qquad |1\rangle \ = \ \frac{1}{\sqrt{d^2-1}}\left(|\hat{1}\rangle - \frac{1}{d}\,|\hat{0}\rangle\right).
\ee
Some explicit examples of evaluations of the topological states are given in Appendix~\ref{sec:examples}.

\section{Recovery of a topological qubit}
\label{sec:topcorrection}

\subsection{Stochastic protocol}
\label{sec:errcorr1}

Let us take a qubit in a generic state, which can always be parameterized in terms of the expansion in an appropriate basis:
\be
\label{logical}
|\psi\rangle \equiv \ \begin{array}{c}
     \begin{tikzpicture}
        \draw[draw=white,double=black,line width=1,rounded corners=2] (0,-0.55) -- (0.7,-0.55) -- (0.7,-0.65) -- (-0.7,-0.65) -- (-0.7,-0.55) -- (0,-0.55);
        \draw[draw=white,double=black,line width=1,rounded corners=2] (0.6,-0.6) -- (0.6,-0.7) -- (-0.6,-0.7) -- (-0.6,-0.5) -- (0.45,-0.5);
        \draw[draw=white,double=black,line width=1,rounded corners=2] (-0.05,-0.5) -- (0.1,-0.5) -- (0.1,-0.7) -- (0.35,-0.7) -- (.35,-0.5) -- (0.6,-0.5) -- (0.6,-0.6);
        \draw[draw=white,double=black,line width=1,rounded corners=2] (-0.45,0) -- (-0.45,-0.7) -- (0,-0.7);
        \draw[draw=white,double=black,line width=1,rounded corners=2] (-0.15,0) -- (-0.15,-0.5) -- (-0.35,-0.5) -- (-0.35,-0.7) -- (-0.55,-0.7) -- (-0.55,-0.6);
        \draw[draw=white,double=black,line width=1,rounded corners=2] (-0.5,-0.5) -- (-0.7,-0.5) -- (-0.7,-0.6) -- (0.25,-0.6) -- (0.25,-0.5) -- (0.15,-0.5) -- (0.15,0);
        \draw[draw=white,double=black,line width=1,rounded corners=2] (0.45,0) -- (0.45,-0.6);
         \fill[olive,opacity=0.8,rounded corners=3] (0,-.2) -- (0.75,-.2) -- (0.75,0.2) -- (-0.75,0.2) -- (-0.75,-.2) -- (0,-.2);
         \foreach \x in {-0.45,-0.15,...,0.75}
           \fill[black] (\x,0) circle (0.075);
     \end{tikzpicture} 
\end{array} \ = \ \alpha\begin{array}{c}
     \begin{tikzpicture}
        \draw[draw=white,double=black,line width=1,rounded corners=2] (-0.45,0) -- (-0.45,-0.6) -- (-0.15,-0.6) -- (-0.15,0);
        \draw[draw=white,double=black,line width=1,rounded corners=2] (0.45,0) -- (0.45,-0.6) -- (0.15,-0.6) -- (0.15,0);
         \fill[olive,opacity=0.8,rounded corners=3] (0,-.2) -- (0.75,-.2) -- (0.75,0.2) -- (-0.75,0.2) -- (-0.75,-.2) -- (0,-.2);
         \foreach \x in {-0.45,-0.15,...,0.75}
           \fill[black] (\x,0) circle (0.075);
     \end{tikzpicture} 
\end{array} 
+ \beta\begin{array}{c}
     \begin{tikzpicture}
        \draw[draw=white,double=black,line width=1,rounded corners=2] (-0.45,0) -- (-0.45,-0.7) -- (0.45,-0.7) -- (0.45,0);
        \draw[draw=white,double=black,line width=1,rounded corners=2] (0.15,0) -- (0.15,-0.5) -- (-0.15,-0.5) -- (-0.15,0);
         \fill[olive,opacity=0.8,rounded corners=3] (0,-.2) -- (0.75,-.2) -- (0.75,0.2) -- (-0.75,0.2) -- (-0.75,-.2) -- (0,-.2);
         \foreach \x in {-0.45,-0.15,...,0.75}
           \fill[black] (\x,0) circle (0.075);
     \end{tikzpicture} 
\end{array} \ = \ {\alpha} |\hat0\rangle + {\beta} |\hat 1\rangle \ = \ \hat{\alpha} |0\rangle + \hat{\beta} |1\rangle\,.
\ee
Here it is convenient to expand the state in the non-orthogonal diagrammatic basis $|\hat0\rangle$ and $|\hat1\rangle$, rather than the orthogonal computational basis $|0\rangle$ and $|1\rangle$. Coefficients $\alpha$ and $\beta$ ($\hat\alpha$ and $\hat\beta$) are by default unknown. Using~(\ref{overlaps}), the expansion coefficients in the two bases are related via
\be
\label{2bases}
\hat{\alpha} \ = \ \alpha d +\beta\,, \qquad \hat\beta \ = \ \beta\sqrt{d^2-1}\,.
\ee

We will encode this qubit in four physical qubits, through the following diagram,
\be
\label{encodingchip}
\begin{array}{c}
     \begin{tikzpicture}
         \fill[gray,opacity=0.5,rounded corners=5] (0,-2) -- (2,-2) -- (2,2) -- (-2,2) -- (-2,-2) -- (0,-2);
         \fill[olive,rounded corners=3] (0,-2.3) -- (1,-2.3) -- (1,-1.7) -- (-1,-1.7) -- (-1,-2.3) -- (0,-2.3);
         \fill[olive,rounded corners=3] (0,2.3) -- (1,2.3) -- (1,1.7) -- (-1,1.7) -- (-1,2.3) -- (0,2.3);
         \fill[olive,rounded corners=3] (2.3,0) -- (2.3,1) -- (1.7,1) -- (1.7,-1) -- (2.3,-1) -- (2.3,0);
         \fill[olive,rounded corners=3] (-2.3,0) -- (-2.3,1) -- (-1.7,1) -- (-1.7,-1) -- (-2.3,-1) -- (-2.3,0);
         \fill[white,rounded corners=3] (0,-.3) -- (1,-.3) -- (1,0.3) -- (-1,0.3) -- (-1,-.3) -- (0,-.3);
         \draw[line width=1.5,rounded corners=3] (-2,0.2) -- (-1.4,0.2) -- (-0.8,0.8) -- (-0.3,0.8);
         \draw[line width=1.5,rounded corners=3] (-0.1,0.8) -- (0.8,0.8) -- (1.4,0.2) -- (2,0.2);
         \draw[line width=1.5,rounded corners=3] (0.2,2) -- (0.2,1.2) -- (0.5,0.9);
         \draw[line width=1.5,rounded corners=3]  (0.7,0.7) -- (1.1,0.3) -- (1.1,-0.4) -- (1.0,-0.5);
         \draw[line width=1.5,rounded corners=3]  (0.8,-0.7) -- (0.2,-1.4) -- (0.2,-2);
         \draw[line width=1.5,rounded corners=3] (-0.6,-2) -- (-0.6,-1.4) -- (-1.4,-0.6) -- (-2,-0.6);
         \draw[line width=1.5,rounded corners=3] (0.6,-2) -- (0.6,-1.4) -- (1.4,-0.6) -- (2,-0.6);
         \draw[line width=1.5,rounded corners=3] (-0.6,2) -- (-0.6,1.4) -- (-1.4,0.6) -- (-2,0.6);
         \draw[line width=1.5,rounded corners=3] (0.6,2) -- (0.6,1.4) -- (1.4,0.6) -- (2,0.6);
         \draw[line width=1.5,rounded corners=3] (-2,-0.2) -- (-1.4,-0.2) -- (-1,-0.6) -- (-0.7,-0.6);
         \draw[line width=1.5,rounded corners=3] (-0.5,-0.6) -- (-0.2,-0.6) -- (-0.2,0);
         \draw[line width=1.5,rounded corners=3] (2,-0.2) -- (1.4,-0.2) -- (1.2,-0.6) -- (0.2,-0.6) -- (0.2,0);
         \draw[line width=1.5,rounded corners=3] (-0.2,-2) -- (-0.2,-1.4) -- (-0.6,-1) -- (-0.6,0);
         \draw[line width=1.5,rounded corners=3] (0.6,0) -- (0.6,0.6) -- (-0.2,0.6) -- (-0.2,2);
         \foreach \x in {-0.6,-0.2,...,1}
           \fill[black] (\x,0) circle (0.075);
           \foreach \x in {-0.6,-0.2,...,1}
           \fill[black] (\x,-2) circle (0.075);
           \foreach \x in {-0.6,-0.2,...,1}
           \fill[black] (\x,2) circle (0.075);
           \foreach \x in {-0.6,-0.2,...,1}
           \fill[black] (-2,\x) circle (0.075);
           \foreach \x in {-0.6,-0.2,...,1}
           \fill[black] (2,\x) circle (0.075);
     \end{tikzpicture} 
\end{array}
\ee
The logical qubit is encoded by filling the cavity in the center, closing the open ends in the form compatible with state~(\ref{logical}). The result is a 3-space with four $S^2$ boundaries, with the information about the original qubit encoded in the connections between the spheres. Each sphere inherits a connection from the logical qubit and shares one connection with each one of the remaining three. Using the basis~(\ref{ONbasis}) and the technique explained in section~\ref{sec:tqubits} one can find the explicit form of the above state, which can be found in Appendix~\ref{sec:coefficients} (see equation~(\ref{coeffs})).

We will attempt the recovery of the original logical qubit in the case when the rightmost physical qubit is lost, and we want to rebuild the logical qubit in the leftmost one. 

Recovering the original qubit essentially means collecting all the connections inherited from it by the spheres in the selected leftmost qubit. Naively, one might think to achieve this through exchanging the endpoints between different available qubits through appropriate nonlocal permutation operations. If one could collect all the connections with the logical qubit in a single physical one then the original state would be recovered by a local permutation, unless the previous step created a topological obstruction, ``tying'' the state to the rest of the system. There are two obstacles however.

First, it is not possible to collect all the original connections, because one of them is presumably lost. We will argue here and in section~\ref{sec:errcorr3} that losing one connection, by itself, does not preclude the information recovery, because of the redundancy, with which the TQFT states are organized. In this section, in particular, we will show that the lost connection can be recovered with a finite probability of success.

\begin{figure} 
    $$
    \begin{array}{c} 
    \includegraphics[width=0.2\linewidth,angle=-90]{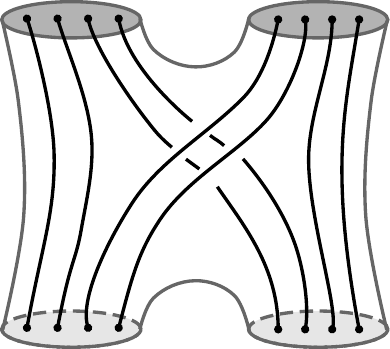}\end{array} \qquad \Longrightarrow \qquad \begin{array}{c}\includegraphics[width=0.2\linewidth,angle=-90]{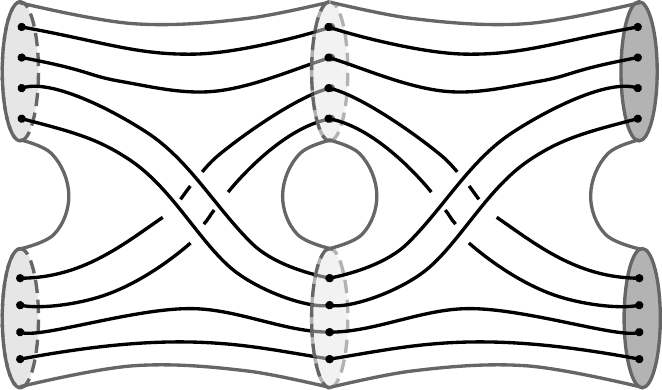}\end{array} \quad \neq \quad 
    \begin{array}{c}\includegraphics[width=0.2\linewidth,angle=-90]{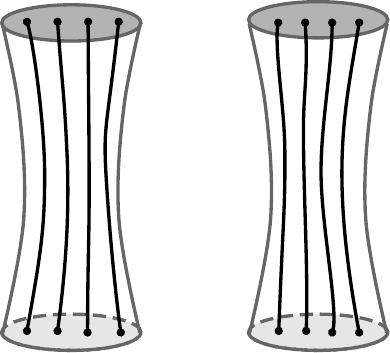}\end{array}
    $$
    \caption{Topological operators acting on tensor products of Hilbert spaces are in general nonunitary: The concatenation of the left operator with its Hermitian conjugate forms the space shown in the center. The result cannot be converted into the identity operator (right) due to the topological defect (hole) in the center.}
    \label{fig:nonunitop}
\end{figure}

Second, it turns out that the topological toolbox does not possess appropriate nonlocal unitary permutations. The main problem is that any topology realizing an operator ${\cal{O}}:\Hc_4\otimes\Hc_4\to \Hc_4\otimes\Hc_4$ is either a combination of a local unitary and a trivial swap of two $\Hc_4$ factors, or a nonunitary operator -- an invertible operator or a projector. Specifically, the operator of the first type just permutes qubits as a whole and acts on them locally, while the one of the second type creates additional 3-dimensional defects (holes) in the topology. Figure~\ref{fig:nonunitop} shows why such operators are in general nonunitary and how holes appear in the topology. 

The last problem can be solved in principle if ${\cal O}$ embeds $\Hc_4\otimes \Hc_4$ in a larger Hilbert space, where nontrivial permutations can act unitarily. We will do this in the next section, by also slightly modifying the encoding of the physical qubit. Note that encoding~(\ref{encodingchip}) is by itself nonunitary, that is corresponds to a nonisometric tensor. The reason for this and for the generic nonunitarity of nonlocal permutation ${\cal{O}}$, such as the one shown in figure~\ref{fig:nonunitop}, is that such operations implicitly incorporate Hilbert spaces $\Hc_n$ with higher number $n$ of marked points on a sphere ($\Hc_8$ for ${\cal{O}}$ and $\Hc_{16}$ for the encoding). Therefore the employed operations first embed lower dimensional Hilbert space in $\Hc_n$ and then project the result on a subspace. The projection occurs because $\Hc_{n_1+n_2}\neq \Hc_{n_1}\otimes\Hc_{n_2}$, which can be seen, for example, from their dimensions~(\ref{Catalan}).

As stated, in this section we will solve the recovery problem in a probabilistic fashion, therefore providing a classical protocol for the recovery. Note that the encoding can be successfully achieved, e.g through entanglement swapping protocol, on 5-qubit chip equivalent to~(\ref{encodingchip}). In such a protocol the logical qubit is teleported to the fifth qubit in the center of ~(\ref{encodingchip}). Possible unitaries affecting the logical qubit in this protocol can be included in the definition of the logical qubit to be dealt later. The appropriate protocol is schematically shown in Appendix~\ref{sec:coefficients}. 

Now we assume that we have managed to successfully encode the logical qubit. Note that the simplest way to collect the available connections of the logical qubit in the target qubit is by closing the outputs of the spheres as the following diagram shows:
\be
\label{projection}
\begin{array}{c}
     \begin{tikzpicture}
         \fill[gray,opacity=0.5,rounded corners=5] (0,-2) -- (2,-2) -- (2,2) -- (-2,2) -- (-2,-2) -- (0,-2);
         \fill[olive,rounded corners=3] (0,-2.3) -- (1,-2.3) -- (1,-1.7) -- (-1,-1.7) -- (-1,-2.3) -- (0,-2.3);
         \fill[olive,rounded corners=3] (0,2.3) -- (1,2.3) -- (1,1.7) -- (-1,1.7) -- (-1,2.3) -- (0,2.3);
         \fill[olive,rounded corners=3] (-2.3,0) -- (-2.3,1) -- (-1.7,1) -- (-1.7,-1) -- (-2.3,-1) -- (-2.3,0);
         \fill[white,rounded corners=3] (0,-.3) -- (1,-.3) -- (1,0.3) -- (-1,0.3) -- (-1,-.3) -- (0,-.3);
         \draw[line width=1.5,rounded corners=3] (-2,0.2) -- (-1.4,0.2) -- (-0.8,0.8) -- (-0.3,0.8);
         \draw[line width=1.5,rounded corners=3] (-0.1,0.8) -- (0.8,0.8) -- (1.4,0.2) -- (2,0.2);
         \draw[line width=1.5,rounded corners=3] (0.2,2) -- (0.2,1.2) -- (0.5,0.9);
         \draw[line width=1.5,rounded corners=3]  (0.7,0.7) -- (1.1,0.3) -- (1.1,-0.4) -- (1.0,-0.5);
         \draw[line width=1.5,rounded corners=3]  (0.8,-0.7) -- (0.2,-1.4) -- (0.2,-2);
         \draw[line width=1.5,rounded corners=3] (-0.6,-2) -- (-0.6,-1.4) -- (-1.4,-0.6) -- (-2,-0.6);
         \draw[line width=1.5,rounded corners=3] (0.6,-2) -- (0.6,-1.4) -- (1.4,-0.6) -- (2,-0.6);
         \draw[line width=1.5,rounded corners=3] (-0.6,2) -- (-0.6,1.4) -- (-1.4,0.6) -- (-2,0.6);
         \draw[line width=1.5,rounded corners=3] (0.6,2) -- (0.6,1.4) -- (1.4,0.6) -- (2,0.6);
         \draw[line width=1.5,rounded corners=3] (-2,-0.2) -- (-1.4,-0.2) -- (-1,-0.6) -- (-0.7,-0.6);
         \draw[line width=1.5,rounded corners=3] (-0.5,-0.6) -- (-0.2,-0.6) -- (-0.2,0);
         \draw[line width=1.5,rounded corners=3] (2,-0.2) -- (1.4,-0.2) -- (1.2,-0.6) -- (0.2,-0.6) -- (0.2,0);
         \draw[line width=1.5,rounded corners=3] (-0.2,-2) -- (-0.2,-1.4) -- (-0.6,-1) -- (-0.6,0);
         \draw[line width=1.5,rounded corners=3] (0.6,0) -- (0.6,0.6) -- (-0.2,0.6) -- (-0.2,2);
         \draw[line width=3,blue]
         (-0.6,2) arc (180:0:0.2);
         \draw[line width=3,blue]
         (-0.6,-2) arc (-180:0:0.2);
         \draw[line width=3,orange]
         (0.2,2) arc (180:0:0.2);
         \draw[line width=3,orange]
         (0.2,-2) arc (-180:0:0.2);
         \foreach \x in {-0.6,-0.2,...,1}
           \fill[black] (\x,0) circle (0.075);
           \foreach \x in {-0.6,-0.2,...,1}
           \fill[black] (\x,-2) circle (0.075);
           \foreach \x in {-0.6,-0.2,...,1}
           \fill[black] (\x,2) circle (0.075);
           \foreach \x in {-0.6,-0.2,...,1}
           \fill[black] (-2,\x) circle (0.075);
     \end{tikzpicture} 
\end{array}
\qquad
\Longrightarrow
\qquad
\begin{array}{c}
     \begin{tikzpicture}
         \fill[gray,opacity=0.5,rounded corners=5] (0,-2) -- (2,-2) -- (2,2) -- (-2,2) -- (-2,-2) -- (0,-2);
         \fill[olive,rounded corners=3] (-2.3,0) -- (-2.3,1) -- (-1.7,1) -- (-1.7,-1) -- (-2.3,-1) -- (-2.3,0);
         \fill[white,rounded corners=3] (0,-.3) -- (1,-.3) -- (1,0.3) -- (-1,0.3) -- (-1,-.3) -- (0,-.3);
         \draw[line width=1.5,rounded corners=3] (-2,0.2) -- (-1.4,0.2) -- (-0.8,0.8) -- (-0.2,0.8);
         \draw[line width=1.5,rounded corners=3] (0.2,0.8) -- (0.8,0.8) -- (1.4,0.2) -- (2,0.2);
         \draw[line width=1.5,rounded corners=3] (-2,-0.6) -- (-1.4,-0.6) -- (-1.,-1.) -- (-0.6,-1) -- (-0.6,0);
         \draw[line width=1.5,rounded corners=3] (-2,0.6) -- (-1.4,0.6) -- (-0.6,1.4) -- (-0.2,1.4) -- (-0.2,1) -- (0.2,0.6) -- (0.6,0.6) -- (0.6,0);
         \draw[line width=1.5,rounded corners=3] (-2,-0.2) -- (-1.4,-0.2) -- (-1.0,-0.6) -- (-0.7,-0.6);
         \draw[line width=1.5,rounded corners=3] (-0.5,-0.6) -- (-0.2,-0.6) -- (-0.2,0);
         \draw[line width=1.5,rounded corners=3] (2,-0.2) -- (1.4,-0.2) -- (1.0,-0.6) -- (0.2,-0.6) -- (0.2,0);
         \foreach \x in {-0.6,-0.2,...,1}
           \fill[black] (\x,0) circle (0.075);
           \foreach \x in {-0.6,-0.2,...,1}
           \fill[black] (-2,\x) circle (0.075);
     \end{tikzpicture} 
\end{array}
\ee
Such an operation is not unitary and can be viewed as a result of a measurement of the state $|0\rangle$ on both the upper, and the lower qubits. After the measurement the system is equivalent to the right diagram above. At the end of this section we will compute the probability of this outcome.

Now assume our measurement of the upper and lower qubits gave $|0\rangle$ for both of them. Note that although the rightmost qubit is lost, the system still inherits connections (correlations) with it. This means that the fate of the lines exiting diagram~(\ref{projection}) eastbound is not known. In other words, the state we are dealing with appears to be mixed.

As the result of the described manipulations the target (leftmost) qubit gains three of the four connections of the original qubit, but the system still appears to be correlated with the lost qubit, although the connection is only through a pair of Wilson lines. The last fact means that the system is in fact uncorrelated with the lost part, that is, it is in a pure state, which is a well-known property of a 3-space that can be disconnected from the rest by a cut along an $S^2$ with at most two marked points~\cite{Witten:1988hf}:\footnote{See an explicit example illustrating this property in Appendix~\ref{sec:examples} (example 3).}
\be
\label{split}
\begin{array}{c}
     \begin{tikzpicture}
         \fill[gray,opacity=0.5,rounded corners=5] (0,-2) -- (2,-2) -- (2,2) -- (-2,2) -- (-2,-2) -- (0,-2);
         \fill[olive,rounded corners=3] (-2.3,0) -- (-2.3,1) -- (-1.7,1) -- (-1.7,-1) -- (-2.3,-1) -- (-2.3,0);
         \fill[olive,rounded corners=3] (1.8,0) -- (1.8,2.1) -- (1.2,2.1) -- (1.2,-2.1) -- (1.8,-2.1) -- (1.8,0);
         \fill[white,rounded corners=3] (0,-.3) -- (1,-.3) -- (1,0.3) -- (-1,0.3) -- (-1,-.3) -- (0,-.3);
         \draw[line width=1.5,rounded corners=3] (-2,0.2) -- (-1.4,0.2) -- (-0.8,0.8) -- (-0.2,0.8);
         \draw[line width=1.5,rounded corners=3] (0.2,0.8) -- (0.8,0.8) -- (1.4,0.2) -- (2,0.2);
         \draw[line width=1.5,rounded corners=3] (-2,-0.6) -- (-1.4,-0.6) -- (-1.,-1.) -- (-0.6,-1) -- (-0.6,0);
         \draw[line width=1.5,rounded corners=3] (-2,0.6) -- (-1.4,0.6) -- (-0.6,1.4) -- (-0.2,1.4) -- (-0.2,1) -- (0.2,0.6) -- (0.6,0.6) -- (0.6,0);
         \draw[line width=1.5,rounded corners=3] (-2,-0.2) -- (-1.4,-0.2) -- (-1.0,-0.6) -- (-0.7,-0.6);
         \draw[line width=1.5,rounded corners=3] (-0.5,-0.6) -- (-0.2,-0.6) -- (-0.2,0);
         \draw[line width=1.5,rounded corners=3] (2,-0.2) -- (1.4,-0.2) -- (1.0,-0.6) -- (0.2,-0.6) -- (0.2,0);
         \foreach \x in {-0.6,-0.2,...,1}
           \fill[black] (\x,0) circle (0.075);
           \foreach \x in {-0.6,-0.2,...,1}
           \fill[black] (-2,\x) circle (0.075);
        \fill[black] (1.5,0.2) circle (0.075);
        \fill[black] (1.5,-0.2) circle (0.075);
     \end{tikzpicture} 
\end{array}
\qquad = \qquad 
\begin{array}{c}
     \begin{tikzpicture}
         \fill[gray,opacity=0.5,rounded corners=5] (0,-2) -- (2,-2) -- (2,2) -- (-2,2) -- (-2,-2) -- (0,-2);
         \fill[olive,rounded corners=3] (-2.3,0) -- (-2.3,1) -- (-1.7,1) -- (-1.7,-1) -- (-2.3,-1) -- (-2.3,0);
         \fill[olive,rounded corners=3] (1.8,0) -- (1.8,2.1) -- (1.2,2.1) -- (1.2,-2.1) -- (1.8,-2.1) -- (1.8,0);
         \fill[white,rounded corners=3] (0,-.3) -- (1,-.3) -- (1,0.3) -- (-1,0.3) -- (-1,-.3) -- (0,-.3);
         \draw[line width=1.5,rounded corners=3] (-2,0.2) -- (-1.4,0.2) -- (-0.8,0.8) -- (-0.2,0.8);
         \draw[line width=1.5,rounded corners=3] (0.2,0.8) -- (0.8,0.8) -- (1.4,0.2) -- (1.5,0.2);
         \draw[line width=1.5,rounded corners=3] (-2,-0.6) -- (-1.4,-0.6) -- (-1.,-1.) -- (-0.6,-1) -- (-0.6,0);
         \draw[line width=1.5,rounded corners=3] (-2,0.6) -- (-1.4,0.6) -- (-0.6,1.4) -- (-0.2,1.4) -- (-0.2,1) -- (0.2,0.6) -- (0.6,0.6) -- (0.6,0);
         \draw[line width=1.5,rounded corners=3] (-2,-0.2) -- (-1.4,-0.2) -- (-1.0,-0.6) -- (-0.7,-0.6);
         \draw[line width=1.5,rounded corners=3] (-0.5,-0.6) -- (-0.2,-0.6) -- (-0.2,0);
         \draw[line width=1.5,rounded corners=3] (1.5,-0.2) -- (1.4,-0.2) -- (1.0,-0.6) -- (0.2,-0.6) -- (0.2,0);
         \foreach \x in {-0.6,-0.2,...,1}
           \fill[black] (\x,0) circle (0.075);
           \foreach \x in {-0.6,-0.2,...,1}
           \fill[black] (-2,\x) circle (0.075);
        \draw[line width=1.5,rounded corners=3] (1.5,0.2) arc (90:-90:0.2);
        \fill[black] (1.5,0.2) circle (0.075);
        \fill[black] (1.5,-0.2) circle (0.075);
        \fill[gray,opacity=0.5,rounded corners=5] (2.4,0) -- (2.4,2) -- (3.6,2) -- (3.6,-2) -- (2.4,-2) -- (2.4,0);
         \fill[olive,rounded corners=3] (3.2,0) -- (3.2,2.1) -- (2.6,2.1) -- (2.6,-2.1) -- (3.2,-2.1) -- (3.2,0);
         \draw[line width=1.5,rounded corners=3] (2.9,0.2) arc (90:270:0.2);
         \draw[line width=1.5,rounded corners=3] (2.9,-0.2) -- (3.6,-0.2);
         \draw[line width=1.5,rounded corners=3] (2.9,0.2) -- (3.6,0.2);
        \fill[black] (2.9,0.2) circle (0.075);
        \fill[black] (2.9,-0.2) circle (0.075);
     \end{tikzpicture} 
\end{array}
\ee
In other words, as the result of these simple manipulations, the target qubit is found in a separable state. This state differs from the original qubit by a braiding of lines that can be corrected by a local operation:
\be
\label{target}
\begin{array}{c}
     \begin{tikzpicture}
        \draw[draw=white,double=black,line width=1,rounded corners=2] (0,-0.55) -- (0.7,-0.55) -- (0.7,-0.65) -- (-0.7,-0.65) -- (-0.7,-0.55) -- (0,-0.55);
        \draw[draw=white,double=black,line width=1,rounded corners=2] (0.6,-0.6) -- (0.6,-0.7) -- (-0.6,-0.7) -- (-0.6,-0.5) -- (0.45,-0.5);
        \draw[draw=white,double=black,line width=1,rounded corners=2] (-0.05,-0.5) -- (0.1,-0.5) -- (0.1,-0.7) -- (0.35,-0.7) -- (.35,-0.5) -- (0.6,-0.5) -- (0.6,-0.6);
        \draw[draw=white,double=black,line width=1,rounded corners=2] (-0.45,0) -- (-0.45,-0.7) -- (0,-0.7);
        \draw[draw=white,double=black,line width=1,rounded corners=2] (-0.15,0) -- (-0.15,-0.5) -- (-0.35,-0.5) -- (-0.35,-0.7) -- (-0.55,-0.7) -- (-0.55,-0.6);
        \draw[draw=white,double=black,line width=1,rounded corners=2] (-0.5,-0.5) -- (-0.7,-0.5) -- (-0.7,-0.6) -- (0.25,-0.6) -- (0.25,-0.5) -- (0.15,-0.5) -- (0.15,0);
        \draw[draw=white,double=black,line width=1,rounded corners=2] (0.45,0) -- (0.45,-0.6);
         \fill[gray,opacity=0.2,rounded corners=3] (0,-.2) -- (0.75,-.2) -- (0.75,0.2) -- (-0.75,0.2) -- (-0.75,-.2) -- (0,-.2);
         \foreach \x in {-0.45,-0.15,...,0.75}
           \fill[black] (\x,0) circle (0.075);
           \draw[draw=white,double=black,line width=1,rounded corners=2] (-0.45,0) -- (-0.45,0.4) -- (-0.15,0.7) -- (-0.15,1.1);
           \draw[draw=white,double=black,line width=1,rounded corners=2] (-0.15,0) -- (-0.15,0.4) -- (-0.45,0.7) -- (-0.45,1.1);
           \draw[draw=white,double=black,line width=1,rounded corners=2] (0.45,0) -- (0.45,0.4) -- (0.15,0.7) -- (0.15,1.1);
           \draw[draw=white,double=black,line width=1,rounded corners=2] (0.15,0) -- (0.15,0.4) -- (0.45,0.7) -- (0.45,1.1);
            \fill[olive,opacity=0.8,rounded corners=3] (0,0.9) -- (0.75,.9) -- (0.75,1.3) -- (-0.75,1.3) -- (-0.75,.9) -- (0,.9);
            \foreach \x in {-0.45,-0.15,...,0.75}
           \fill[black] (\x,1.1) circle (0.075);
     \end{tikzpicture} 
\end{array}
\qquad \Longrightarrow \qquad
\begin{array}{c}
     \begin{tikzpicture}
        \draw[draw=white,double=black,line width=1,rounded corners=2] (0,-0.55) -- (0.7,-0.55) -- (0.7,-0.65) -- (-0.7,-0.65) -- (-0.7,-0.55) -- (0,-0.55);
        \draw[draw=white,double=black,line width=1,rounded corners=2] (0.6,-0.6) -- (0.6,-0.7) -- (-0.6,-0.7) -- (-0.6,-0.5) -- (0.45,-0.5);
        \draw[draw=white,double=black,line width=1,rounded corners=2] (-0.05,-0.5) -- (0.1,-0.5) -- (0.1,-0.7) -- (0.35,-0.7) -- (.35,-0.5) -- (0.6,-0.5) -- (0.6,-0.6);
        \draw[draw=white,double=black,line width=1,rounded corners=2] (-0.45,0) -- (-0.45,-0.7) -- (0,-0.7);
        \draw[draw=white,double=black,line width=1,rounded corners=2] (-0.15,0) -- (-0.15,-0.5) -- (-0.35,-0.5) -- (-0.35,-0.7) -- (-0.55,-0.7) -- (-0.55,-0.6);
        \draw[draw=white,double=black,line width=1,rounded corners=2] (-0.5,-0.5) -- (-0.7,-0.5) -- (-0.7,-0.6) -- (0.25,-0.6) -- (0.25,-0.5) -- (0.15,-0.5) -- (0.15,0);
        \draw[draw=white,double=black,line width=1,rounded corners=2] (0.45,0) -- (0.45,-0.6);
         \fill[olive,opacity=0.8,rounded corners=3] (0,-.2) -- (0.75,-.2) -- (0.75,0.2) -- (-0.75,0.2) -- (-0.75,-.2) -- (0,-.2);
         \foreach \x in {-0.45,-0.15,...,0.75}
           \fill[black] (\x,0) circle (0.075);
     \end{tikzpicture} 
\end{array}
\ee
This completes the diagrammatic construction.

Now we would like to describe the same protocol in the computational basis. Qubit~(\ref{logical}) is encoded in the 16-dimensional Hilbert space, so it has the form
\be
\label{physical2}
|\psi\rangle \ \to\ |\Psi\rangle \ = \ a_{0000}|0000\rangle + a_{0001} |0001\rangle + a_{1000} |1000\rangle + a_{1001} |1001\rangle + \ldots
\ee
Here we have only kept the relevant coefficients, while the full set can be found in Appendix~\ref{sec:coefficients}. The first digit of the state label corresponds to the target qubit, the second and the third one to the top and to the bottom, respectively, and the last one is the lost qubit. Since we are projecting on the $|0\rangle$ state of the top and bottom qubits, we do not need the remaining coefficients here. The values of the necessary coefficients are
\begin{eqnarray}
a_{0000} & = &d^{-3}(\alpha d+\beta)\,, \label{a0000}\\
a_{0001} & = & d^{-3}\sqrt{d^2-1}(\alpha d+\beta)\,, \\
a_{1000} & = & d^{-3}\sqrt{d^2-1}\beta\,,\\
a_{1001} & = & d^{-3}(d^2-1)\beta\,. \label{a1001}
\end{eqnarray}

It is not hard to check that the result of the projection is the separable state
\be
\label{result}
\langle00|\Psi\rangle \ = \ \left((\alpha d+\beta)|0\rangle +\beta\sqrt{d^2-1}|1\rangle\right)\otimes \left(d^{-3}|0\rangle+d^{-3}\sqrt{d^2-1}|1\rangle\right).
\ee
All the information about the original state is contained in the first qubit, which is, in fact, in the same state as the original qubit:
\be
\label{target2}
(\alpha d+\beta)|0\rangle +\beta\sqrt{d^2-1}|1\rangle \ = \ \alpha |\hat{0}\rangle + \beta|\hat{1}\rangle\,,
\ee
in accordance with~(\ref{logical}) and~(\ref{2bases}).

\begin{figure}[htb]
    \centering
    \includegraphics[width=0.65\linewidth]{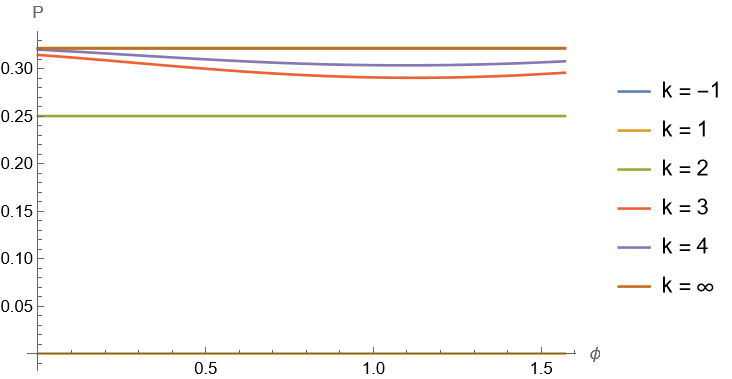}
    \caption{Probability of a successfull recovery of the logical qubit from state~(\ref{encodingchip}) for fixed integer values of parameter $k$~(\ref{params}). The probabilities for $k = -1$ (blue) and large $k$ (brown) overlap. Here, we use a parameterization $\hat{\alpha}=\sin\phi$ and $\hat{\beta}=\cos\phi$~(\ref{logical}).}
    \label{Figure 1}
\end{figure}

The same trick as in~(\ref{split}) can be applied if the top is measured in the $|0\rangle$ state and the bottom -- in the $|1\rangle$, or vice versa. The target qubit becomes separable also in this outcome. This happens because two of the four lines that connect to the lost qubit in the left diagram of~(\ref{encodingchip}) are forced to form a spin 1 state, so there will be only one possible way to form a singlet with the remaining pair of spin 1/2 lines. This simple prediction of separability can be easily checked from the explicit form in the computational basis. The result of the measurement in these cases is
\be
\langle01|\Psi\rangle \ = \ A^8\left((\alpha d+\beta)|0\rangle -\frac{\beta}{\sqrt{d^2-1}}|1\rangle\right)\otimes \left(d^{-3}\sqrt{d^2-1}|0\rangle-d^{-3}|1\rangle\right)\,,
\ee
\be
\langle10|\Psi\rangle \ = \ A^{-8}\left({\beta}{\sqrt{d^2-1}}|0\rangle +(A^8\alpha d-\beta)|1\rangle\right)\otimes \left(d^{-3}|0\rangle-\frac{d^{-3}}{\sqrt{d^2-1}}|1\rangle\right)\,.
\ee

Again, in the last case the information about the original state $|\psi\rangle$ (parameters $\alpha$ and $\beta$) remains accessible, but to recover the state itself, some knowledge about the state is necessary. In other words, there is no general, $\alpha$ and $\beta$ independent unitary that rotates the target qubit to $|\psi\rangle$.

If the top and bottom qubits are both measured in the $|1\rangle$ state, the resulting state is nonseparable.

\begin{figure}[htb]
    \centering
    \includegraphics[width=0.65\linewidth]{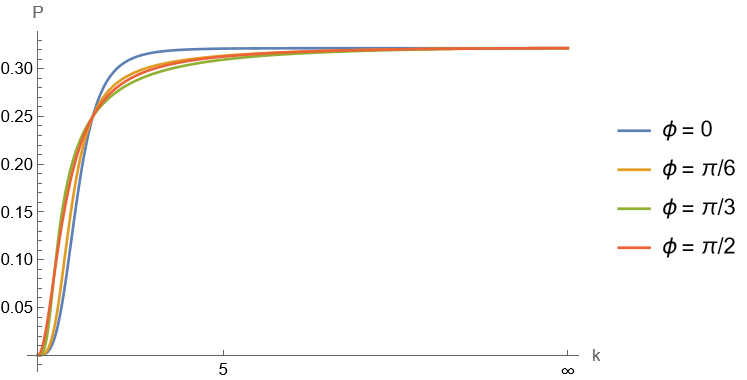}
    \caption{Probability for fixed values of the parameter $\phi$. As $k \rightarrow \infty$, the probability $\mathbb{P} \rightarrow {9}/{28}$.}
    \label{Figure 2}
\end{figure}

The explicit parameterization can be used to compute the probabilities, for example, of measuring both upper and lower qubits in the $|0\rangle$ state. In terms of the coefficients of $|\Psi\rangle$~(\ref{physical2}), the unnormalized post-measurement density matrix reads
\be
\rho^{\prime} \ = \  \sum\limits_{\substack{i,l}} \sum\limits_{\substack{m,p}}a_{i00l} a_{m00p}^* |i00l\rangle \langle m00p|\,. 
\ee
The probability is given by the trace
\be
\mathbb{P} = \frac{\text{Tr}_{i,l}(\rho^{\prime})}{\text{Tr}(\rho)} = \left({\sum\limits_{\substack{i,l}} |a_{i00l}|^2}\right)/\left({\sum\limits_{\substack{i,l,m,p}} |a_{ilmp}|^2}\right)\,.
\ee
Using the full set of coefficients~(\ref{coeffs}), calculated in the appendix, one can compute this probability, for example, for real choice of the parameters $\alpha$ and $\beta$ ($\hat{\alpha}$ and $\hat{\beta}$). We show the result of these calculations for different values of the parameter in figure~\ref{Figure 1}. As one infers from the figure the probability is almost independent of the details of the logical qubit and stays slightly above 30\%, with the exception of $k=1$ (degenerate qubit) and $k=2$ (25\% probability).

One can see the $k$ dependence of the probability for a fixed choice of the parameterization of $\hat{\alpha}$ e $\hat{\beta}$ in figure~\ref{Figure 2}. The maximum value $\mathbb{P} = {9}/{28}$ is attained in the classical limit of Chern-Simons $k\to \infty$ and coincides with another special (non-Chern-Simons) case $k=-1$.

\subsection{Braiding protocol}
\label{sec:errcorr2}

In the previous section, we considered a setup, in which a logical qubit was encoded in a direct product of four topological single-qubit Hilbert spaces $\Hc_4$. Now we are going to consider another version of the protocol with physical qubits realized in a larger Hilbert space, which does not involve projectors. More precisely, instead of thinking of four physical qubits as associated with four punctures on four distinct spheres, we will consider the situation with sixteen punctures on a single $S^2$. The naive dimension of such Hilbert space is $2^{16}=65536$, but TQFT constrains the spins to be in a singlet state, which corresponds to the maximal dimension given by the 8th Catalan number $C_8=1430$.

The encoding transformation maps the four punctures of the logical qubit to the sixteen punctures of the physical qubits, separated into four groups. Again we encode the logical qubit in such a way that the information is spread evenly among the physical ones. This can be illustrated by the following diagram:
\be
\begin{array}{c}
     \begin{tikzpicture}
        \fill[olive,opacity=0.9,rounded corners=5] (0,-2.3) -- (2.3,-2.3) -- (2.3,2.3) -- (-2.3,2.3) -- (-2.3,-2.3) -- (0,-2.3);
         \fill[gray,opacity=0.7,rounded corners=5] (0,-1.7) -- (1.7,-1.7) -- (1.7,1.7) -- (-1.7,1.7) -- (-1.7,-1.7) -- (0,-1.7);
         \fill[olive,rounded corners=3] (0,-2.3) -- (1,-2.3) -- (1,-1.7) -- (-1,-1.7) -- (-1,-2.3) -- (0,-2.3);
         \fill[olive,rounded corners=3] (0,2.3) -- (1,2.3) -- (1,1.7) -- (-1,1.7) -- (-1,2.3) -- (0,2.3);
         \fill[olive,rounded corners=3] (2.3,0) -- (2.3,1) -- (1.7,1) -- (1.7,-1) -- (2.3,-1) -- (2.3,0);
         \fill[olive,rounded corners=3] (-2.3,0) -- (-2.3,1) -- (-1.7,1) -- (-1.7,-1) -- (-2.3,-1) -- (-2.3,0);
         \fill[white,rounded corners=3] (0,-.3) -- (1,-.3) -- (1,0.3) -- (-1,0.3) -- (-1,-.3) -- (0,-.3);
         \draw[line width=1.5,rounded corners=3] (-2,0.2) -- (-1.4,0.2) -- (-0.8,0.8) -- (-0.3,0.8);
         \draw[line width=1.5,rounded corners=3] (-0.1,0.8) -- (0.8,0.8) -- (1.4,0.2) -- (2,0.2);
         \draw[line width=1.5,rounded corners=3] (0.2,2) -- (0.2,1.2) -- (0.5,0.9);
         \draw[line width=1.5,rounded corners=3]  (0.7,0.7) -- (1.1,0.3) -- (1.1,-0.4) -- (1.0,-0.5);
         \draw[line width=1.5,rounded corners=3]  (0.8,-0.7) -- (0.2,-1.4) -- (0.2,-2);
         \draw[line width=1.5,rounded corners=3] (-0.6,-2) -- (-0.6,-1.4) -- (-1.4,-0.6) -- (-2,-0.6);
         \draw[line width=1.5,rounded corners=3] (0.6,-2) -- (0.6,-1.4) -- (1.4,-0.6) -- (2,-0.6);
         \draw[line width=1.5,rounded corners=3] (-0.6,2) -- (-0.6,1.4) -- (-1.4,0.6) -- (-2,0.6);
         \draw[line width=1.5,rounded corners=3] (0.6,2) -- (0.6,1.4) -- (1.4,0.6) -- (2,0.6);
         \draw[line width=1.5,rounded corners=3] (-2,-0.2) -- (-1.4,-0.2) -- (-1,-0.6) -- (-0.7,-0.6);
         \draw[line width=1.5,rounded corners=3] (-0.5,-0.6) -- (-0.2,-0.6) -- (-0.2,0);
         \draw[line width=1.5,rounded corners=3] (2,-0.2) -- (1.4,-0.2) -- (1.2,-0.6) -- (0.2,-0.6) -- (0.2,0);
         \draw[line width=1.5,rounded corners=3] (-0.2,-2) -- (-0.2,-1.4) -- (-0.6,-1) -- (-0.6,0);
         \draw[line width=1.5,rounded corners=3] (0.6,0) -- (0.6,0.6) -- (-0.2,0.6) -- (-0.2,2);
         \foreach \x in {-0.6,-0.2,...,1}
           \fill[black] (\x,0) circle (0.075);
           \foreach \x in {-0.6,-0.2,...,1}
           \fill[black] (\x,-2) circle (0.075);
           \foreach \x in {-0.6,-0.2,...,1}
           \fill[black] (\x,2) circle (0.075);
           \foreach \x in {-0.6,-0.2,...,1}
           \fill[black] (-2,\x) circle (0.075);
           \foreach \x in {-0.6,-0.2,...,1}
           \fill[black] (2,\x) circle (0.075);
     \end{tikzpicture} 
\end{array}\,,
\ee
which is the same bulk topology as before, but with a single connected edge.

Let us assume that we have lost access to the rightmost quadruple of punctures. In other words, we assume that the information is corrupted locally, affecting a small region of the sphere, whose characteristic radius is smaller than the characteristic distance between the regions. Then by application of braiding gates on the remaining punctures, we can transform the above diagram into
\be
\begin{array}{c}
     \begin{tikzpicture}
        \fill[olive,opacity=0.9,rounded corners=5] (0,-2.3) -- (2.3,-2.3) -- (2.3,2.3) -- (-2.3,2.3) -- (-2.3,-2.3) -- (0,-2.3);
         \fill[gray,opacity=0.7,rounded corners=5] (0,-1.7) -- (1.7,-1.7) -- (1.7,1.7) -- (-1.7,1.7) -- (-1.7,-1.7) -- (0,-1.7);
         \fill[olive,rounded corners=3] (0,-2.3) -- (1,-2.3) -- (1,-1.7) -- (-1,-1.7) -- (-1,-2.3) -- (0,-2.3);
         \fill[olive,rounded corners=3] (0,2.3) -- (1,2.3) -- (1,1.7) -- (-1,1.7) -- (-1,2.3) -- (0,2.3);
         \fill[olive,rounded corners=3] (2.3,0) -- (2.3,1) -- (1.7,1) -- (1.7,-1) -- (2.3,-1) -- (2.3,0);
         \fill[olive,rounded corners=3] (-2.3,0) -- (-2.3,1) -- (-1.7,1) -- (-1.7,-1) -- (-2.3,-1) -- (-2.3,0);
         \fill[white,rounded corners=3] (0,-.3) -- (1,-.3) -- (1,0.3) -- (-1,0.3) -- (-1,-.3) -- (0,-.3);
         \draw[orange,line width=1.5,rounded corners=3] (-2,0.6) -- (-1.4,0.6) -- (-1.2,0.8) -- (0.8,0.8) -- (1.4,0.2) -- (2,0.2);
         \draw[line width=1.5,rounded corners=3] (0.2,2) -- (0.2,1.2) -- (0.5,0.9);
         \draw[line width=1.5,rounded corners=3]  (0.7,0.7) -- (1.1,0.3) -- (1.1,-0.4) -- (1.0,-0.5);
         \draw[line width=1.5,rounded corners=3]  (0.8,-0.7) -- (0.2,-1.4) -- (0.2,-2);
         \draw[line width=1.5,rounded corners=3] (-0.6,-2) -- (-0.6,-1.4) -- (-0.2,-1.4) -- (-0.2,-2);
         \draw[line width=1.5,rounded corners=3] (0.6,-2) -- (0.6,-1.4) -- (1.4,-0.6) -- (2,-0.6);
         \draw[line width=1.5,rounded corners=3] (-0.6,2) -- (-0.6,1.4) -- (-0.2,1.4) -- (-0.2,2);
         \draw[line width=1.5,rounded corners=3] (0.6,2) -- (0.6,1.4) -- (1.4,0.6) -- (2,0.6);
         \draw[line width=1.5,rounded corners=3] (-2,-0.2) -- (-1.4,-0.2) -- (-1,-0.6) -- (-0.2,-0.6) -- (-0.2,0);
         \draw[line width=1.5,rounded corners=3] (-0.6,-0.5)  -- (-0.6,0);
         \draw[orange,line width=1.5,rounded corners=3] (2,-0.2) -- (1.4,-0.2) -- (1.2,-0.6) -- (0.2,-0.6) -- (0.2,0);
         \draw[line width=1.5,rounded corners=3] (-2,-0.6) -- (-1.4,-0.6) -- (-1.0,-1.) -- (-0.6,-1) -- (-0.6,-0.7);
         \draw[line width=1.5,rounded corners=3] (0.6,0) -- (0.6,0.6) -- (-1.0,0.6) -- (-1.4,0.2) -- (-2,0.2);
         \foreach \x in {-0.6,-0.2,...,1}
           \fill[black] (\x,0) circle (0.075);
           \foreach \x in {-0.6,-0.2,...,1}
           \fill[black] (\x,-2) circle (0.075);
           \foreach \x in {-0.6,-0.2,...,1}
           \fill[black] (\x,2) circle (0.075);
           \foreach \x in {-0.6,-0.2,...,1}
           \fill[black] (-2,\x) circle (0.075);
           \foreach \x in {-0.6,-0.2,...,1}
           \fill[black] (2,\x) circle (0.075);
           \fill[white,rounded corners=.3] (2.3,0) -- (2.3,1) -- (1.7,1) -- (1.7,-1) -- (2.3,-1) -- (2.3,0);
     \end{tikzpicture} 
\end{array}
\ee
In the case of a single $S^2$ braiding is a unitary operation.

It looks like we can recover most of the information about the original logical qubit in one (leftmost) of the regions encoding topological qubits since in the last diagram the leftmost group of punctures holds three of the four connections to the logical qubit. This group is connected with its complement in the full Hilbert space by only two lines (orange), which implies that its reduced density matrix has rank two. It will be shown that possession of three punctures of the logical qubit is enough to recover the full information about it.

As a first step let us view the diagram in the following way:
\be
\label{16state0}
\begin{array}{c}
     \begin{tikzpicture}
        \fill[olive,opacity=0.9,rounded corners=5] (0,-2.3) -- (2.3,-2.3) -- (2.3,2.3) -- (-2.3,2.3) -- (-2.3,-2.3) -- (0,-2.3);
         \fill[gray,opacity=0.7,rounded corners=5] (0,-1.7) -- (1.7,-1.7) -- (1.7,1.7) -- (-1.7,1.7) -- (-1.7,-1.7) -- (0,-1.7);
         \fill[olive,rounded corners=3] (0,-2.3) -- (1,-2.3) -- (1,-1.7) -- (-1,-1.7) -- (-1,-2.3) -- (0,-2.3);
         \fill[olive,rounded corners=3] (0,2.3) -- (1,2.3) -- (1,1.7) -- (-1,1.7) -- (-1,2.3) -- (0,2.3);
         \fill[olive,rounded corners=3] (2.3,0) -- (2.3,1) -- (1.7,1) -- (1.7,-1) -- (2.3,-1) -- (2.3,0);
         \fill[olive,rounded corners=3] (-2.3,0) -- (-2.3,1) -- (-1.7,1) -- (-1.7,-1) -- (-2.3,-1) -- (-2.3,0);
         \fill[white,rounded corners=3] (0,-.3) -- (1,-.3) -- (1,0.3) -- (-1,0.3) -- (-1,-.3) -- (0,-.3);
         \draw[line width=1.5,rounded corners=3] (-2,0.6) -- (-1.4,0.6) -- (-1.2,0.8) -- (0.8,0.8) -- (1.4,0.2) -- (2,0.2);
         \draw[line width=1.5,rounded corners=3] (0.2,2) -- (0.2,1.2) -- (0.5,0.9);
         \draw[line width=1.5,rounded corners=3]  (0.7,0.7) -- (1.1,0.3) -- (1.1,-0.4) -- (1.0,-0.5);
         \draw[line width=1.5,rounded corners=3]  (0.8,-0.7) -- (0.2,-1.4) -- (0.2,-2);
         \draw[line width=1.5,rounded corners=3] (-0.6,-2) -- (-0.6,-1.4) -- (-0.2,-1.4) -- (-0.2,-2);
         \draw[line width=1.5,rounded corners=3] (0.6,-2) -- (0.6,-1.4) -- (1.4,-0.6) -- (2,-0.6);
         \draw[line width=1.5,rounded corners=3] (-0.6,2) -- (-0.6,1.4) -- (-0.2,1.4) -- (-0.2,2);
         \draw[line width=1.5,rounded corners=3] (0.6,2) -- (0.6,1.4) -- (1.4,0.6) -- (2,0.6);
         \draw[orange,line width=1.5,rounded corners=3] (-2,-0.2) -- (-1.4,-0.2) -- (-1,-0.6) -- (-0.2,-0.6) -- (-0.2,0);
         \draw[orange,line width=1.5,rounded corners=3] (-0.6,-0.5)  -- (-0.6,0);
         \draw[orange,line width=1.5,rounded corners=3] (2,-0.2) -- (1.4,-0.2) -- (1.2,-0.6) -- (0.2,-0.6) -- (0.2,0);
         \draw[orange,line width=1.5,rounded corners=3] (-2,-0.6) -- (-1.4,-0.6) -- (-1.0,-1.) -- (-0.6,-1) -- (-0.6,-0.7);
         \draw[orange,line width=1.5,rounded corners=3] (0.6,0) -- (0.6,0.6) -- (-1.0,0.6) -- (-1.4,0.2) -- (-2,0.2);
         \foreach \x in {-0.6,-0.2,...,1}
           \fill[black] (\x,0) circle (0.075);
           \foreach \x in {-0.6,-0.2,...,1}
           \fill[black] (\x,-2) circle (0.075);
           \foreach \x in {-0.6,-0.2,...,1}
           \fill[black] (\x,2) circle (0.075);
           \foreach \x in {-0.6,-0.2,...,1}
           \fill[black] (-2,\x) circle (0.075);
           \foreach \x in {-0.6,-0.2,...,1}
           \fill[black] (2,\x) circle (0.075);
           \fill[white,rounded corners=.3] (2.3,0) -- (2.3,1) -- (1.7,1) -- (1.7,-1) -- (2.3,-1) -- (2.3,0);
           \draw[orange,dashed] (2.5,-0.2) -- (1.7,-0.2);
           \fill[black] (2.5,-0.2) circle (0.075);
     \end{tikzpicture} 
\end{array}
\ee
The state described by the diagram has the structure
\be
\label{16state}
|\Psi_4\rangle \otimes |\Phi_{12}\rangle\,,
\ee
where $|\Psi_4\rangle$ is the state of four spins connected by the orange lines in the last diagram. Importantly, it is separable from the state $|\Phi_{12}\rangle$ of the remaining twelve spins. This is guaranteed by the fact that four spins of the state $|\Psi_4\rangle$ on the boundary are connected exclusively by the orange lines, so they are in a singlet state. If we focus on $|\Psi_4\rangle$, one can access three of the four spins, while only one line is connected to the spin in the corrupted region. A single line carries no physical information. In the meantime, the remaining three lines are sufficient to distill the information on one of the three spins. In the next section we will show this by explicitly constructing the transformation recovering the original qubit using a (pseudo) spin representation of the state.

\section{Information storage in the topological qubit}
\label{sec:errcorr3}

In this section, we will construct a three-spin transformation on a state of four spins in a topological qubit state, which distills the state of the topological qubit in a single spin. For this, we will use a pseudounitary representation of the topological qubits~\cite{Kauffman:2019top,Padmanabhan:2020obt,Melnikov:2024ddj}. The main reason is that this particular representation has some nice properties, in particular, it acts on tensor products of two-dimensional vector spaces -- the spins. Note that quantum mechanics of Chern-Simons states, as defined in section~\ref{sec:tqubits}, is unitary, so the pseudounitary representation here is equivalent to a different unitary representation of a smaller dimension~\cite{Kauffman:2013bh,Melnikov:2020uck}, which does not distinguish individual spins. The pseudounitary representation is itself unitary for specific values of $k$. For these values, we will find a unitary transformation recovering the logical qubit on one of the spins. One can think of this situation as a physical realization of topological qubits with spins.

\subsection{Pseudounitary representation}
\label{sec:punitary}

Rules of quantum mechanics sketched in section~\ref{sec:tqubits} and the action of braiding operators on the respective states can be realized explicitly as a matrix representation. Let us consider the following matrix
\be
\label{Rmatrix}
R \ = \ \left(
\begin{array}{cccc}
    A & & &   \\
     & A-A^{-3} & A^{-1} & \\
     & A^{-1} & 0 & \\
     &  &  & A
\end{array}
\right).
\ee
This matrix is the R-matrix of $A_3^{(1)}$ generalized Toda system~\cite{Jimbo:1985ua,Turaev:1988eb}. It acts on a tensor product $V\otimes V$ of two two-dimensional linear spaces and can be used to build the generators of the braid group:
\begin{eqnarray*}
    b_1 & = & R\otimes {\mathbb{I}}_2\otimes {\mathbb{I}}_2\otimes \cdots\,,\\
    b_2 & = & {\mathbb{I}}_2\otimes R\otimes {\mathbb{I}}_2\otimes\cdots\,, \\
    b_3 & = & {\mathbb{I}}_2\otimes {\mathbb{I}}_2\otimes R\otimes \cdots\,, \\
    & \cdots &
\end{eqnarray*}
Here ${\mathbb{I}}_2$ is a $2\times 2$ identity matrix.

Matrix~(\ref{Rmatrix}) represents the left-hand side of the skein relation~(\ref{skein}). The inverse braid generator is given by the inverse of the R-matrix and by the left-hand-side diagram of~(\ref{skein}) rotated 90 degrees. In the right-hand side of~(\ref{skein}) the first diagram can be thought as of an identity matrix. Then the second diagram must be a generator of the Temperley-Lieb algebra. Indeed, one can check that
\be
U \ \equiv \quad \begin{tikzpicture}[baseline=0]
\draw[thick] (-0.1,-0.5) -- (-0.1,-0.3) arc (180:0:0.2) -- (0.3,-0.5);
\draw[thick] (-0.1,0.5) -- (-0.1,0.3) arc (-180:0:0.2) -- (0.3,0.5);
\end{tikzpicture} \quad = \ A\, R - A^2\,{\mathbb{I}}_4 \ =    \ \left(
\begin{array}{cccc}
    0 & & &   \\
     & -A^{-2} & 1 & \\
     & 1 & - A^2 & \\
     &  &  & 0
\end{array}
\right).
\ee
and the generators, obtained from $U$ in the same way as the braid generators from matrix $R$, satisfy the Temperley-Lieb relations. (Here $\mathbb{I}_4$ is a $4\times4$ identity matrix.) In particular,
\be
U^2 \ = \ d\, U\,.
\ee
Note that one can simultaneously rotate all the diagrams in~(\ref{skein}) by 90 degrees and one will get another form of the skein relation, now for the inverse R-matrix.

Now we would like to have states~(\ref{basis}) in this representation. Those can be chosen (nonuniquely) as
\begin{eqnarray}
\langle \hat{0}| & = & (|\hat{0}\rangle)^{\rm T} \ = \ \left(0,0,0,0,0,-A^{-2},1,0,0,1,-A^2,0,0,0,0,0\right), \label{mbasis0}\\
\langle \hat{1}| & = & (|\hat{1}\rangle)^{\rm T} \ = \ \left(0,0,0,-A^{-2},0,1,0,0,0,0,1,0,-A^2,0,0,0\right).\label{mbasis1}
\end{eqnarray}
Note that the conjugated vector is computed as a simple transposition. This happens because representation~(\ref{Rmatrix}) is pseudounitary. In particular,
\be
R^{-1} \ = \ \Sigma^{\otimes 2} R^\dagger(\Sigma^{\otimes 2})^\dagger\,,
\ee
where 
\be
\Sigma \ = \ \left(\begin{array}{cc}
    0 & 1 \\
    1 & 0
\end{array}\right)
\ee
and $\Sigma^{\otimes n}$ is the $n^{\rm th}$ tensor power of $\Sigma$. Consequently,
\begin{eqnarray}
\langle \hat{0}| & = & (|\hat{0}\rangle^\ast)^{\rm T}\Sigma^{\otimes 4} \ = \ (|\hat{0}\rangle)^{\rm T}\,, \\ 
\langle \hat{1}| & = & (|\hat{1}\rangle^\ast)^{\rm T}\Sigma^{\otimes 4} \ = \ (|\hat{1}\rangle)^{\rm T}\,.
\end{eqnarray}
Since $A^\ast=A^{-1}$ the conjugated vectors are equivalent to the transposed ones.

It is not hard to check that the pseudounitary representation is also unitary for special values of $A$, namely, when $A=\pm 1$ or $A=\pm i$. This means that in this case the qubit states can be thought as of being composed of four physical spins (rather than pseudospins). However, in the unitary case matrix~(\ref{Rmatrix}) becomes a simple permutation matrix (SWAP gate), which makes it a less interesting quantum operation.

Vectors~(\ref{mbasis0}) and~(\ref{mbasis1}) and their tensor products span the code subspace of the physical Hilbert space of spins. One can check that overlaps~(\ref{overlaps}) are correctly reproduced as well as other diagrammatic properties, and the Jones polynomials of links obtained by the procedure outlined in section~\ref{sec:tqubits} are computed as matrix elements of braids and tangles constructed from the generators of the braid group and Temperley-Lieb algebra. See~\cite{Melnikov:2024ddj} for more details.

For completeness, we also mention the calculation of Markov's trace~\cite{Turaev:1988eb}. As is well known, Jones polynomials and similar topological invariants of knots can be obtained by computing a trace of a braid. The special trace respecting the ambient isotopy of knots and links is called Markov's trace. In the current representation, it can be computed as a normal trace of the braid matrix on $n$ strands with the insertion of the $n^{\rm th}$ tensor power of matrix\footnote{Here notation $q^{H}$ comes from the connection of R matrix and quantum groups. See~\cite{Turaev:1988eb}.}
\be
q^H \ = \ \left(
\begin{array}{cc}
 -A^2 &   \\
     & -A^{-2}  \\
\end{array}
\right), \qquad q^H:\, V\ \to \ V\,.
\ee
Namely,
\be
\Tr_{\rm M}X \ = \ \Tr \left((q^H)^{\otimes n}X\right)\,.
\ee

One can check the trace gives the same results as the calculation of the invariants as matrix elements in the basis generated by~(\ref{mbasis0}) and~(\ref{mbasis1}).

\subsection{Recovery transformation}
\label{sec:encoding}

A (topo)-logical qubit can be expanded in the diagrammatic basis~(\ref{basis}), or equivalently~(\ref{mbasis0}) and~(\ref{mbasis1}). This state encodes the logical qubit in four physical qubits, where we now refer to spins as physical qubits. It does this in such a way that if one qubit is lost the information can still be recovered on the remaining three physical qubits. The precise encoding maps state~(\ref{logical}) to the following state of four qubits (here unnormalized):
\begin{multline}
\label{4spinencoding}
|\Psi\rangle \ = \ |0\rangle\otimes |\Phi\rangle + |1\rangle\otimes |\Phi^\star\rangle
\\ = \ |0\rangle\left(B|011\rangle + (bB+cC)|101\rangle+C|110\rangle \right) + |1\rangle\left(B^\ast|100\rangle + (bB+cC)^\ast|010\rangle+C^\ast|001\rangle \right)\,,    
\end{multline}
where
\begin{eqnarray}
B \ =\  - A^2\beta\,, && C \ = \ \alpha\,, \\
b = -A^{-2}\,, && c \ = \ -A^{2}\,. 
\end{eqnarray}
We will assume that $\alpha$ and $\beta$ are real and that it is the first qubit that is lost or compromised. Note that from the point of view of the discussion in section~\ref{sec:errcorr2} state $|\Psi\rangle$ is the state ${|\Psi_4\rangle}$ of four spins in a large 16-spin system~(\ref{16state0}) and~(\ref{16state}). Note also that $B$ and $C$ depend on $\alpha$ and $\beta$, while $b$ and $c$ do not. 

Let us initially forget that state~(\ref{4spinencoding}) comes from a nonunitary representation and consider it as a normal state in the Hilbert space of four spins. Then the following two vectors are orthogonal to 3-qubit states $|\Phi\rangle$ and $|\Phi^\star\rangle$ respectively:
\begin{eqnarray}
    |\Phi_0\rangle & = &  \frac{1}{\sqrt{1+|b|^2+|c|^2}}\left(-b^\ast|011\rangle + |101\rangle-c^\ast|110\rangle\right)\,, \\
    |\Phi_0^\star\rangle & = & \frac{1}{\sqrt{1+|b|^2+|c|^2}}\left(-b|100\rangle +|010\rangle-c|001\rangle\right)\,. 
\end{eqnarray}

In the 3-qubit space, there is an orthonormal basis given by $|\Phi_0\rangle$, $|\Phi_0^\star\rangle$, $|000\rangle$, $|111\rangle$ and the following four vectors
\begin{eqnarray}
    |\Phi_+\rangle & = &  \frac{1}{\sqrt{(|b|^2+|c|^2)^2+|b|^2+|c|^2}}\left(-b^\ast|011\rangle - (|b|^2+|c|^2)|101\rangle-c^\ast|110\rangle\right)\,, \\
    |\Phi_+^\star\rangle & = & \frac{1}{\sqrt{(|b|^2+|c|^2)^2+|b|^2+|c|^2}}\left(-b|100\rangle - (|b|^2+|c|^2)|010\rangle-c|001\rangle\right)\,, \\
    |\Phi_-\rangle & = &  \frac{1}{\sqrt{|b|^2+|c|^2}}\left(c|011\rangle -b|110\rangle\right) \,, \\
    |\Phi_-^\star\rangle & = & \frac{1}{\sqrt{|b|^2+|c|^2}}\left(c^\ast|100\rangle -b^\ast|001\rangle\right)\,. 
\end{eqnarray}

Consider a unitary matrix
\begin{multline}
\label{correction}
U \ = \ |00\tilde 0\rangle\langle\Phi_+|+|11\tilde 0\rangle\langle\Phi_+^\star|+|00\tilde 1\rangle\langle\Phi_-|+|11\tilde 1\rangle\langle\Phi_-^\star| +
\\ +|01\tilde 0\rangle\langle\Phi_0|+ |10\tilde 0\rangle\langle\Phi_0^\star| + |01\tilde 1\rangle\langle 000| + |10\tilde 1\rangle\langle 111|,
\end{multline}
where $|\tilde 0\rangle$ and $|\tilde 1\rangle$ are two orthogonal, yet unspecified basis vectors of a single qubit. 

Note that by construction, the terms in the second line of~(\ref{correction}) annihilate vectors $|\Phi\rangle$ and $|\Phi^\ast\rangle$, so that operator $\mathbb{I}_2\otimes U$, when applied on $|\Psi\rangle$, yields the following:
\begin{multline}
\mathbb{I}_2\otimes U|\Psi\rangle \ = \ -\frac{(bB+cC)(|b|^2+|c|^2+1)}{\sqrt{(|b|^2+|c|^2)^2+|b|^2+|c|^2}}|000\tilde 0\rangle - \frac{(bB+cC)^\ast(|b|^2+|c|^2+1)}{\sqrt{(|b|^2+|c|^2)^2+|b|^2+|c|^2}}|111\tilde 0\rangle \\
 +  \frac{(c^\ast B-b^\ast C)}{\sqrt{|b|^2+|c|^2}}|000\tilde 1\rangle+ \frac{(c^\ast B-b^\ast C)^\ast}{\sqrt{|b|^2+|c|^2}}|111\tilde 1\rangle \label{CorrState}\\
 =  |000\rangle\left(-\frac{3(bB+cC)}{\sqrt{6}}|\tilde 0\rangle+\frac{(c^\ast B-b^\ast C)}{\sqrt{2}}|\tilde 1\rangle\right) + |111\rangle\left(-\frac{3(bB+cC)^\ast}{\sqrt{6}}|\tilde 0\rangle+\frac{(c^\ast B-b^\ast C)^\ast}{\sqrt{2}}|\tilde 1\rangle\right).
\end{multline}
In the last line, we simplified the coefficients taking into account that $b$ and $c$ are just phases.

To recover state $|\psi\rangle$~(\ref{logical}) on the last qubit one needs to properly choose the basis $|\tilde 0\rangle, |\tilde 1\rangle $. It is not hard to check that
\begin{eqnarray}
|\tilde 0\rangle & = & -\sqrt{\frac{3}{2}}|0\rangle - \sqrt{\frac{1}{2}}|1\rangle\,,\\
|\tilde 1\rangle & = & -\sqrt{\frac{1}{2}}|0\rangle + \sqrt{\frac{3}{2}}|1\rangle
\end{eqnarray}
produce
\be
\mathbb{I}_2\otimes U|\Psi\rangle \ = \ |000\rangle\left((\beta-2A^2\alpha))|0\rangle+\sqrt{3}\beta|1\rangle\right) + |111\rangle\left((\beta-2A^{-2}\alpha))|0\rangle+\sqrt{3}\beta|1\rangle\right),
\ee
which coincides with $|\psi\rangle$ written in the orthonormal basis, (\ref{target2}) if $A=\pm 1$ and $d=-2$, or $A=\pm i$ and $d=2$. These are exactly the two cases for which the pseudounitary representation is also unitary.  

In other words, for unitary representations, that is when the topological qubit is physically constructed with real spins, the result is
\be
\mathbb{I}_2\otimes U|\Psi\rangle \ = \ \left(|000\rangle+|111\rangle\right)\otimes\left((\alpha d+\beta)|0\rangle+\sqrt{d^2-1}\beta|1\rangle\right) \ = \ \left(|000\rangle+|111\rangle\right)\otimes|\psi\rangle\,,
\ee
and the original qubit is recovered on the last spin.

\section{Discussion}
\label{sec:conclusions}

In this work we explored the properties of topological quantum field theory (TQFT) states, focusing on their potential for information storage and error correction. As a specific resource, we used quantum states of $SU(2)$ Chern-Simons theory in three-dimensional spaces with two-sphere boundaries. For a given dimension of the Hilbert space such states provide a basis and come with naturally built-in unitary operations -- nonabelian permutation of punctures and the associated braiding of Wilson lines.

Chern-Simons realization with fundamental Wilson lines requires a minimum of four punctures on the associated two-sphere. This results in intrinsic nonlocality of the information storage by the topological qubits. This is a general hallmark of qubits realized in the topological states of matter. One specific feature of this type of storage was discussed in this work: if one loses access to or trust in part of the physical storage, which is not too large, the remaining part still contains the full knowledge of the stored information. 

In the case of a single qubit, encoded by four punctures (spins) on a sphere, any three punctures have access to the full information. We showed this by constructing an explicit transformation that distills the original encoded logical qubit on a physical spin (puncture) operating on only three of four spins of the qubit. In the meantime, no natural topological operation can help extract the information from only one puncture. Manipulations on a single puncture can at most create a local knotting of a single Wilson line, which is just equivalent to an overall phase. 

Nonlocality of the information storage and quantum correlations made explicit by the topological presentation of quantum states, suggest general principles for information manipulation, for example in constructing quantum error-correcting codes on multiple qubits. Wilson line presentation of correlations in a single qubit should be generalized to the multiqubit case, following: $(i)$ the information is spread uniformly across the physical system so that every sufficiently large part of it contains enough to correct possible errors, and $(ii)$ highly nonlocal entangled state is used for the encoding, that is a state maximally and uniformly connected, so that one is able to distill the information efficiently. 

The two simple protocols for error correction considered in this work showed that these general principles work, but also highlighted technical issues related to the intrinsic properties of the topological realization of quantum mechanics. In these protocols, the information, in general, remains accessible in subsystems, although it might not be easily extractable using natural topological tools. In the case of the first protocol, additional efforts were necessary to deterministically encode the logical qubit in a space, which is a naive direct product of four-qubit spaces -- a consequence of the zero net spin constraint in the topological Hilbert spaces. For the same reason, the set of available nonlocal topological operators was essentially restricted to nonunitary operators. As a result, the information was also only available probabilistically.

The issues of the first protocol were solved in the second one, which encoded the logical qubit in a much larger Hilbert space. Such a space is more natural in physical systems, such as quantum Hall anyons, but it is more complex due to the same zero net spin constraint. Information recovery was demonstrated by using an explicit pseudoanyonic representation for qubits. The pseudanyonic representation is unitary for special values of parameter $A$ of the topological model, which includes the semiclassical limit of Chern-Simons $A=1$. In this case, the pseudoanyons are physical anyons, or spins, collectively composing the topological qubits. The information was recovered on a single physical spin by applying a unitary transformation on three uncompromised physical spins.

It would also be interesting to adapt or generalize the protocols discussed here to more specific error correction tasks. Here we only used the general principles $(i)$ and $(ii)$ and quite general realizations of quantum states to highlight error-correcting properties of the TQFT codes. One may hope that supplementing this approach with more specific principles and technologies of quantum error correction can produce useful and novel protocols. 

\paragraph{Acknowledgments.} This work was partially supported by Simons Foundation award number 1023171-RC. The authors also gratefully acknowledge the support of the Brazilian National Council for Scientific and Technological Development (CNPq), through grants numbered 307295/2020-6, 308580/2022-2 and 404274/2023-4, Serrapilheira Institute, through grant number Serra R-2012-38185 (DM) and Grant No. Serra-1708-15763 (RC and DP), Brazilian Ministry of Education and the CAPES Foundation (MN and LP). DM would also like to thank the Isaac Newton Institute for Mathematical Sciences, Cambridge, for support and hospitality during the programme "Black holes: bridges between number theory and holographic quantum information" (EPSRC grant number EP/R014604/1), where part of the work on this paper was undertaken.

\begin{appendix}
   
    \section{Examples of the TQFT representation of quantum states}
\label{sec:examples}

In order to provide some intuition regarding the diagrammatic representation of states, let us consider a few examples:

\begin{enumerate}

\item We will start with the  maximally entangled Bell state \( |\phi^+\rangle \). The reduced density matrix of this state is proportional to the identity matrix \( \mathbb{I}_2 \). One can easily show that states sharing this property are appropriately represented by the following diagram:
\be
\label{phiplus}
|\phi^{+}\rangle \ \equiv 
   \frac{1}{\sqrt{2}}\begin{array}{c}
        \begin{tikzpicture}
            \newcommand{\y}{2}
         \newcommand{\z}{0.37}
         \draw[draw=white,double=black,line width=1,rounded corners=2] (-0.45,0-\z) -- (-0.45,-0.7-\z) -- (0.45+\y,-0.7-\z) -- (0.45+\y,0-\z);
         \draw[draw=white,double=black,line width=1,rounded corners=2] (-0.15,0-\z) -- (-0.15,-0.6-\z) -- (0.15+\y,-0.6-\z) -- (0.15+\y,0-\z);
         \draw[draw=white,double=black,line width=1,rounded corners=2] (0.15,-\z) -- (0.15,-0.5-\z) -- (-0.15+\y,-0.5-\z) -- (-0.15+\y,0-\z);
         \draw[draw=white,double=black,line width=1,rounded corners=2] (0.45,0-\z) -- (0.45,-0.4-\z) -- (-0.45+\y,-0.4-\z) -- (-0.45+\y,0-\z);
        
         \fill[olive,opacity=0.8,rounded corners=3] (0,-.2-\z) -- (0.75,-.2-\z) -- (0.75,0.2-\z) -- (-0.75,0.2-\z) -- (-0.75,-.2-\z) -- (0,-.2-\z);
         \foreach \x in {-0.45,-0.15,...,0.75}
         \fill[black] (\x,-\z) circle (0.075);
         \fill[olive,opacity=0.8,rounded corners=3] (0+\y,-.2-\z) -- (0.75+\y,-.2-\z) -- (0.75+\y,0.2-\z) -- (-0.75+\y,0.2-\z) -- (-0.75+\y,-.2-\z) -- (0+\y,-.2-\z); 
         \foreach \x in {-0.45,-0.15,...,0.75}
         \fill[black] (\x+\y,-\z) circle (0.075);
        \end{tikzpicture}
    \end{array}.
\ee

This representation intuitively illustrates the strong correlation between the two subsystems, say Alice and Bob, with all punctures of one 2-sphere connected to the punctures of the other. The density matrix is by definition $\rho = |\phi\rangle \langle\phi|$, which, in the diagrammatic presentation is obtained by doubling the state diagram. The partial trace can be computed simply by connecting the lines of two 2-spheres corresponding to the same subsystem. This yields a diagram equivalent to the identity matrix:
\be
     \rho_{A} \ = \  {\color{orange}\text{Tr}_{B}}(\rho_{AB})  \ \equiv \  
\frac{1}{2}

\ = \ \frac{1}{2} \ \begin{array}{c}
     \raisebox{-.8ex}{\begin{tikzpicture}
     \newcommand{\z}{0.37}
            \draw[draw=white,double=black,line width=1,rounded corners=2] 
            (0.45,-\z)--(0.45,-2);
            \draw[draw=white,double=black,line width=1,rounded corners=2] 
            (0.15,-\z)--(0.15,-2);        
            \draw[draw=white,double=black,line width=1,rounded corners=2]
            (-0.15,-\z)--(-0.15,-2);      
            \draw[draw=white,double=black,line width=1,rounded corners=2]
            (-0.45,-\z)--(-0.45,-2); 
            
            \fill[olive,opacity=0.8,rounded corners=3] (0,-.2-\z) -- (0.75,-.2-\z) -- (0.75,0.2-\z) -- (-0.75,0.2-\z) -- (-0.75,-.2-\z) -- (0,-.2-\z);
            \foreach \x in {-0.45,-0.15,...,0.75}
            \fill[black] (\x,-\z) circle (0.075);
            
            \fill[gray,opacity=0.5,rounded corners=3] (0,-1.8) -- (0.75,-1.8) -- (0.75,-2.2) -- (-0.75,-2.2) -- (-0.75,-1.8) -- (0,-1.8);
            \foreach \x in {-0.45,-0.15,...,0.75}
            \fill[black] (\x,-2) circle (0.075);
     \end{tikzpicture}} 
\end{array}
\quad \equiv \ \frac{1}{2} \,\mathbb{I}_2.
\ee

It is also possible to test the fidelity of the diagram~(\ref{phiplus}) in the description of the state $|\phi^+\rangle$ by starting from a generic state in the computational basis
\be
|\phi\rangle = \phi_1|00\rangle + \phi_2|01\rangle + \phi_3|10\rangle + \phi_4|11\rangle\,.
\ee

To calculate the coefficients of this expansion, we need the diagrammatic representation of the elements of this basis. From definitions~(\ref{basis}) and~(\ref{ONbasis}) one obtains
\be
\langle 0| = \dfrac{1}{d}
\begin{array}{c}
     \raisebox{2ex}{\begin{tikzpicture}
         \fill[gray,opacity=0.5,rounded corners=3] (0,-.2) -- (0.75,-.2) -- (0.75,0.2) -- (-0.75,0.2) -- (-0.75,-.2) -- (0,-.2);
         \foreach \x in {-0.45,-0.15,...,0.75}
         \fill[black] (\x,0) circle (0.075);
         
         \draw[draw=white,double=black,line width=1,rounded corners=2] (-0.45,0) -- (-0.45,0.6) -- (-0.15,0.6) -- (-0.15,0);
         \draw[draw=white,double=black,line width=1,rounded corners=2] (0.45,0) -- (0.45,0.6) -- (0.15,0.6) -- (0.15,0);
     \end{tikzpicture}}
\end{array} \ , \qquad \langle1| \ = \ \frac{1}{\sqrt{d^2-1}}
\left(\begin{array}{c}
     \raisebox{2ex}{\begin{tikzpicture}
         \fill[gray,opacity=0.5,rounded corners=3] (0,-.2) -- (0.75,-.2) -- (0.75,0.2) -- (-0.75,0.2) -- (-0.75,-.2) -- (0,-.2);
         \foreach \x in {-0.45,-0.15,...,0.75}
         \fill[black] (\x,0) circle (0.075);

         \draw[draw=white,double=black,line width=1,rounded corners=2] (-0.45,0) -- (-0.45,0.6) -- (0.45,0.6) -- (0.45,0);
        \draw[draw=white,double=black,line width=1,rounded corners=2] (0.15,0) -- (0.15,0.4) -- (-0.15,0.4) -- (-0.15,0);
     \end{tikzpicture}}
     \end{array}
     
     \ - \ \frac{1}{d}\begin{array}{c}
    \raisebox{2ex}{\begin{tikzpicture}
    
         \fill[gray,opacity=0.5,rounded corners=3] (0,-.2) -- (0.75,-.2) -- (0.75,0.2) -- (-0.75,0.2) -- (-0.75,-.2) -- (0,-.2);
         \foreach \x in {-0.45,-0.15,...,0.75}
         \fill[black] (\x,0) circle (0.075);

         \draw[draw=white,double=black,line width=1,rounded corners=2] (-0.45,0) -- (-0.45,0.6) -- (-0.15,0.6) -- (-0.15,0);
         \draw[draw=white,double=black,line width=1,rounded corners=2] (0.45,0) -- (0.45,0.6) -- (0.15,0.6) -- (0.15,0);
     \end{tikzpicture}}
     
\end{array} \right).
\ee

Consequently,
\be
\langle 00|\phi\rangle \ \equiv 
    \frac{1}{d^2}\begin{array}{c}
        \begin{tikzpicture}
            \newcommand{\y}{2}
         \newcommand{\z}{0.37}
         
         \draw[draw=white,double=black,line width=1,rounded corners=2] (-0.45,0-\z) -- (-0.45,-0.7-\z) -- (0.45+\y,-0.7-\z) -- (0.45+\y,0-\z);
         \draw[draw=white,double=black,line width=1,rounded corners=2] (-0.15,0-\z) -- (-0.15,-0.6-\z) -- (0.15+\y,-0.6-\z) -- (0.15+\y,0-\z);
         \draw[draw=white,double=black,line width=1,rounded corners=2] (0.15,-\z) -- (0.15,-0.5-\z) -- (-0.15+\y,-0.5-\z) -- (-0.15+\y,0-\z);
         \draw[draw=white,double=black,line width=1,rounded corners=2] (0.45,0-\z) -- (0.45,-0.4-\z) -- (-0.45+\y,-0.4-\z) -- (-0.45+\y,0-\z);
        
         \fill[olive,opacity=0.8,rounded corners=3] (0,-.2-\z) -- (0.75,-.2-\z) -- (0.75,0.2-\z) -- (-0.75,0.2-\z) -- (-0.75,-.2-\z) -- (0,-.2-\z);
         \foreach \x in {-0.45,-0.15,...,0.75}
         \fill[black] (\x,-\z) circle (0.075);
         
         \fill[olive,opacity=0.8,rounded corners=3] (0+\y,-.2-\z) -- (0.75+\y,-.2-\z) -- (0.75+\y,0.2-\z) -- (-0.75+\y,0.2-\z) -- (-0.75+\y,-.2-\z) -- (0+\y,-.2-\z); 
         \foreach \x in {-0.45,-0.15,...,0.75}
         \fill[black] (\x+\y,-\z) circle (0.075);
         
         \draw[draw=white,double=black,line width=1,rounded corners=2] (-0.45,-\z) -- (-0.45,0.6-\z) -- (-0.15,0.6-\z) -- (-0.15,-\z);
         \draw[draw=white,double=black,line width=1,rounded corners=2] (0.45,-\z) -- (0.45,0.6-\z) -- (0.15,0.6-\z) -- (0.15,-\z);
         
         \draw[draw=white,double=black,line width=1,rounded corners=2] (-0.45+\y,-\z) -- (-0.45+\y,0.6-\z) -- (-0.15+\y,0.6-\z) -- (-0.15+\y,-\z);
         \draw[draw=white,double=black,line width=1,rounded corners=2] (0.15+\y,-\z) -- (0.15+\y,0.6-\z) -- (0.45+\y,0.6-\z) -- (0.45+\y,-\z);
        \end{tikzpicture} 
    \end{array} \ = \ \frac{d^2}{d^2} \ =  \ 1\,,
\ee

\be
\begin{aligned}

\langle 01|\phi\rangle & \equiv 
    \frac{1}{\sqrt{d^2-1}} 
    \left(\ \frac{1}{d}\begin{array}{c}
        \begin{tikzpicture}
            \newcommand{\y}{2}
         \newcommand{\z}{0.37}
         
         \draw[draw=white,double=black,line width=1,rounded corners=2] (-0.45,0-\z) -- (-0.45,-0.7-\z) -- (0.45+\y,-0.7-\z) -- (0.45+\y,0-\z);
         \draw[draw=white,double=black,line width=1,rounded corners=2] (-0.15,0-\z) -- (-0.15,-0.6-\z) -- (0.15+\y,-0.6-\z) -- (0.15+\y,0-\z);
         \draw[draw=white,double=black,line width=1,rounded corners=2] (0.15,-\z) -- (0.15,-0.5-\z) -- (-0.15+\y,-0.5-\z) -- (-0.15+\y,0-\z);
         \draw[draw=white,double=black,line width=1,rounded corners=2] (0.45,0-\z) -- (0.45,-0.4-\z) -- (-0.45+\y,-0.4-\z) -- (-0.45+\y,0-\z);
        
         \fill[olive,opacity=0.8,rounded corners=3] (0,-.2-\z) -- (0.75,-.2-\z) -- (0.75,0.2-\z) -- (-0.75,0.2-\z) -- (-0.75,-.2-\z) -- (0,-.2-\z);
         \foreach \x in {-0.45,-0.15,...,0.75}
         \fill[black] (\x,-\z) circle (0.075);
         
         \fill[olive,opacity=0.8,rounded corners=3] (0+\y,-.2-\z) -- (0.75+\y,-.2-\z) -- (0.75+\y,0.2-\z) -- (-0.75+\y,0.2-\z) -- (-0.75+\y,-.2-\z) -- (0+\y,-.2-\z); 
         \foreach \x in {-0.45,-0.15,...,0.75}
         \fill[black] (\x+\y,-\z) circle (0.075);

         \draw[draw=white,double=black,line width=1,rounded corners=2] (-0.45,-\z) -- (-0.45,0.6-\z) -- (-0.15,0.6-\z) -- (-0.15,-\z);
         \draw[draw=white,double=black,line width=1,rounded corners=2] (0.45,-\z) -- (0.45,0.6-\z) -- (0.15,0.6-\z) -- (0.15,-\z);
         
         \draw[draw=white,double=black,line width=1,rounded corners=2] (-0.45+\y,-\z) -- (-0.45+\y,0.6-\z) -- (0.45+\y,0.6-\z) -- (0.45+\y,-\z);
         \draw[draw=white,double=black,line width=1,rounded corners=2] (-0.15+\y,-\z) -- (-0.15+\y,0.5-\z) -- (0.15+\y,0.5-\z) -- (0.15+\y,-\z);
         
        \end{tikzpicture} 
    \end{array}\right), \\ \\
    & = \frac{1}{\sqrt{d^2-1}}\left(\frac{d}{d} - \frac{d^2}{d^2}\right) \ = \  \langle 10|\phi\rangle  \ = \ 0\,,
\end{aligned}
\ee
\be
\begin{aligned}
\langle 11|\phi\rangle \ & \equiv 
    \frac{1}{d^2-1}\left(
    \begin{array}{c}
        \begin{tikzpicture}
            \newcommand{\y}{2}
         \newcommand{\z}{0.37}
         
         \draw[draw=white,double=black,line width=1,rounded corners=2] (-0.45,0-\z) -- (-0.45,-0.7-\z) -- (0.45+\y,-0.7-\z) -- (0.45+\y,0-\z);
         \draw[draw=white,double=black,line width=1,rounded corners=2] (-0.15,0-\z) -- (-0.15,-0.6-\z) -- (0.15+\y,-0.6-\z) -- (0.15+\y,0-\z);
         \draw[draw=white,double=black,line width=1,rounded corners=2] (0.15,-\z) -- (0.15,-0.5-\z) -- (-0.15+\y,-0.5-\z) -- (-0.15+\y,0-\z);
         \draw[draw=white,double=black,line width=1,rounded corners=2] (0.45,0-\z) -- (0.45,-0.4-\z) -- (-0.45+\y,-0.4-\z) -- (-0.45+\y,0-\z);
        
         \fill[olive,opacity=0.8,rounded corners=3] (0,-.2-\z) -- (0.75,-.2-\z) -- (0.75,0.2-\z) -- (-0.75,0.2-\z) -- (-0.75,-.2-\z) -- (0,-.2-\z);
         \foreach \x in {-0.45,-0.15,...,0.75}
         \fill[black] (\x,-\z) circle (0.075);
         
         \fill[olive,opacity=0.8,rounded corners=3] (0+\y,-.2-\z) -- (0.75+\y,-.2-\z) -- (0.75+\y,0.2-\z) -- (-0.75+\y,0.2-\z) -- (-0.75+\y,-.2-\z) -- (0+\y,-.2-\z); 
         \foreach \x in {-0.45,-0.15,...,0.75}
         \fill[black] (\x+\y,-\z) circle (0.075);

         \draw[draw=white,double=black,line width=1,rounded corners=2] (-0.45,-\z) -- (-0.45,0.6-\z) -- (0.45,0.6-\z) -- (0.45,-\z);
         \draw[draw=white,double=black,line width=1,rounded corners=2] (-0.15,-\z) -- (-0.15,0.5-\z) -- (0.15,0.5-\z) -- (0.15,-\z);
         
         \draw[draw=white,double=black,line width=1,rounded corners=2] (-0.45+\y,-\z) -- (-0.45+\y,0.6-\z) -- (0.45+\y,0.6-\z) -- (0.45+\y,-\z);
         \draw[draw=white,double=black,line width=1,rounded corners=2] (-0.15+\y,-\z) -- (-0.15+\y,0.5-\z) -- (0.15+\y,0.5-\z) -- (0.15+\y,-\z);
         
        \end{tikzpicture} 
    \end{array}\ - \
    \frac{1}{d}
    \begin{array}{c}
        \begin{tikzpicture}
         \newcommand{\y}{2}
         \newcommand{\z}{0.37}
         
         \draw[draw=white,double=black,line width=1,rounded corners=2] (-0.45,0-\z) -- (-0.45,-0.7-\z) -- (0.45+\y,-0.7-\z) -- (0.45+\y,0-\z);
         \draw[draw=white,double=black,line width=1,rounded corners=2] (-0.15,0-\z) -- (-0.15,-0.6-\z) -- (0.15+\y,-0.6-\z) -- (0.15+\y,0-\z);
         \draw[draw=white,double=black,line width=1,rounded corners=2] (0.15,-\z) -- (0.15,-0.5-\z) -- (-0.15+\y,-0.5-\z) -- (-0.15+\y,0-\z);
         \draw[draw=white,double=black,line width=1,rounded corners=2] (0.45,0-\z) -- (0.45,-0.4-\z) -- (-0.45+\y,-0.4-\z) -- (-0.45+\y,0-\z);
        
         \fill[olive,opacity=0.8,rounded corners=3] (0,-.2-\z) -- (0.75,-.2-\z) -- (0.75,0.2-\z) -- (-0.75,0.2-\z) -- (-0.75,-.2-\z) -- (0,-.2-\z);
         \foreach \x in {-0.45,-0.15,...,0.75}
         \fill[black] (\x,-\z) circle (0.075);
         
         \fill[olive,opacity=0.8,rounded corners=3] (0+\y,-.2-\z) -- (0.75+\y,-.2-\z) -- (0.75+\y,0.2-\z) -- (-0.75+\y,0.2-\z) -- (-0.75+\y,-.2-\z) -- (0+\y,-.2-\z); 
         \foreach \x in {-0.45,-0.15,...,0.75}
         \fill[black] (\x+\y,-\z) circle (0.075);

         \draw[draw=white,double=black,line width=1,rounded corners=2] (-0.45,-\z) -- (-0.45,0.6-\z) -- (0.45,0.6-\z) -- (0.45,-\z);
         \draw[draw=white,double=black,line width=1,rounded corners=2] (-0.15,-\z) -- (-0.15,0.5-\z) -- (0.15,0.5-\z) -- (0.15,-\z);
         
         \draw[draw=white,double=black,line width=1,rounded corners=2] (-0.45+\y,-\z) -- (-0.45+\y,0.6-\z) -- (-0.15+\y,0.6-\z) -- (-0.15+\y,-\z);
         \draw[draw=white,double=black,line width=1,rounded corners=2] (0.15+\y,-\z) -- (0.15+\y,0.6-\z) -- (0.45+\y,0.6-\z) -- (0.45+\y,-\z);
         
        \end{tikzpicture}     
    \end{array} \right. \\
    & \hspace{1.15cm} - \
    \frac{1}{d}
    \begin{array}{c}
        \begin{tikzpicture}
            \newcommand{\y}{2}
         \newcommand{\z}{0.37}
         
         \draw[draw=white,double=black,line width=1,rounded corners=2] (-0.45,0-\z) -- (-0.45,-0.7-\z) -- (0.45+\y,-0.7-\z) -- (0.45+\y,0-\z);
         \draw[draw=white,double=black,line width=1,rounded corners=2] (-0.15,0-\z) -- (-0.15,-0.6-\z) -- (0.15+\y,-0.6-\z) -- (0.15+\y,0-\z);
         \draw[draw=white,double=black,line width=1,rounded corners=2] (0.15,-\z) -- (0.15,-0.5-\z) -- (-0.15+\y,-0.5-\z) -- (-0.15+\y,0-\z);
         \draw[draw=white,double=black,line width=1,rounded corners=2] (0.45,0-\z) -- (0.45,-0.4-\z) -- (-0.45+\y,-0.4-\z) -- (-0.45+\y,0-\z);
        
         \fill[olive,opacity=0.8,rounded corners=3] (0,-.2-\z) -- (0.75,-.2-\z) -- (0.75,0.2-\z) -- (-0.75,0.2-\z) -- (-0.75,-.2-\z) -- (0,-.2-\z);
         \foreach \x in {-0.45,-0.15,...,0.75}
         \fill[black] (\x,-\z) circle (0.075);
         
         \fill[olive,opacity=0.8,rounded corners=3] (0+\y,-.2-\z) -- (0.75+\y,-.2-\z) -- (0.75+\y,0.2-\z) -- (-0.75+\y,0.2-\z) -- (-0.75+\y,-.2-\z) -- (0+\y,-.2-\z); 
         \foreach \x in {-0.45,-0.15,...,0.75}
         \fill[black] (\x+\y,-\z) circle (0.075);

         \draw[draw=white,double=black,line width=1,rounded corners=2] (-0.45,-\z) -- (-0.45,0.6-\z) -- (-0.15,0.6-\z) -- (-0.15,-\z);
         \draw[draw=white,double=black,line width=1,rounded corners=2] (0.45,-\z) -- (0.45,0.6-\z) -- (0.15,0.6-\z) -- (0.15,-\z);
         
         \draw[draw=white,double=black,line width=1,rounded corners=2] (-0.45+\y,-\z) -- (-0.45+\y,0.6-\z) -- (0.45+\y,0.6-\z) -- (0.45+\y,-\z);
         \draw[draw=white,double=black,line width=1,rounded corners=2] (-0.15+\y,-\z) -- (-0.15+\y,0.5-\z) -- (0.15+\y,0.5-\z) -- (0.15+\y,-\z);
         
        \end{tikzpicture} 
    \end{array}\ + \
    \frac{1}{d^2}
    \left.\begin{array}{c}
        \begin{tikzpicture}
            \newcommand{\y}{2}
         \newcommand{\z}{0.37}
         
         \draw[draw=white,double=black,line width=1,rounded corners=2] (-0.45,0-\z) -- (-0.45,-0.7-\z) -- (0.45+\y,-0.7-\z) -- (0.45+\y,0-\z);
         \draw[draw=white,double=black,line width=1,rounded corners=2] (-0.15,0-\z) -- (-0.15,-0.6-\z) -- (0.15+\y,-0.6-\z) -- (0.15+\y,0-\z);
         \draw[draw=white,double=black,line width=1,rounded corners=2] (0.15,-\z) -- (0.15,-0.5-\z) -- (-0.15+\y,-0.5-\z) -- (-0.15+\y,0-\z);
         \draw[draw=white,double=black,line width=1,rounded corners=2] (0.45,0-\z) -- (0.45,-0.4-\z) -- (-0.45+\y,-0.4-\z) -- (-0.45+\y,0-\z);
        
         \fill[olive,opacity=0.8,rounded corners=3] (0,-.2-\z) -- (0.75,-.2-\z) -- (0.75,0.2-\z) -- (-0.75,0.2-\z) -- (-0.75,-.2-\z) -- (0,-.2-\z);
         \foreach \x in {-0.45,-0.15,...,0.75}
         \fill[black] (\x,-\z) circle (0.075);
         
         \fill[olive,opacity=0.8,rounded corners=3] (0+\y,-.2-\z) -- (0.75+\y,-.2-\z) -- (0.75+\y,0.2-\z) -- (-0.75+\y,0.2-\z) -- (-0.75+\y,-.2-\z) -- (0+\y,-.2-\z); 
         \foreach \x in {-0.45,-0.15,...,0.75}
         \fill[black] (\x+\y,-\z) circle (0.075);

         \draw[draw=white,double=black,line width=1,rounded corners=2] (-0.45,-\z) -- (-0.45,0.6-\z) -- (-0.15,0.6-\z) -- (-0.15,-\z);
         \draw[draw=white,double=black,line width=1,rounded corners=2] (0.45,-\z) -- (0.45,0.6-\z) -- (0.15,0.6-\z) -- (0.15,-\z);
         
         \draw[draw=white,double=black,line width=1,rounded corners=2] (-0.45+\y,-\z) -- (-0.45+\y,0.6-\z) -- (-0.15+\y,0.6-\z) -- (-0.15+\y,-\z);
         \draw[draw=white,double=black,line width=1,rounded corners=2] (0.15+\y,-\z) -- (0.15+\y,0.6-\z) -- (0.45+\y,0.6-\z) -- (0.45+\y,-\z);
         
        \end{tikzpicture}
    \end{array}\right) \\ \\
    & = \ \frac{1}{d^2-1}\left(d^2 -\frac{2d}{d} +\frac{d^2}{d^2} \right) \ = \ 1\,.
\end{aligned}
\ee

We then obtained $\phi_1 = \phi_4 = 1$ and $\phi_2 = \phi_3 = 0$. Therefore,

\be
|\phi\rangle = \frac{|00\rangle + |11\rangle}{\sqrt{2}} = |\phi^+\rangle .
\ee

\item Now consider a state described by the diagram
\be
   \begin{array}{c}
        \begin{tikzpicture}
            \newcommand{\y}{2}
         \newcommand{\z}{0.37}
         \draw[draw=white,double=black,line width=1,rounded corners=2] (-0.45,0-\z) -- (-0.45,-0.7-\z) -- (-0.15,-0.7-\z) -- (-0.15,0-\z);
         \draw[draw=white,double=black,line width=1,rounded corners=2] (0.15,0-\z) -- (0.15,-0.7-\z) -- (0.45,-0.7-\z) -- (0.45,0-\z);
         \draw[draw=white,double=black,line width=1,rounded corners=2] (-0.45+\y,0-\z) -- (-0.45+\y,-0.7-\z) -- (-0.15+\y,-0.7-\z) -- (-0.15+\y,0-\z);
         \draw[draw=white,double=black,line width=1,rounded corners=2] (0.15+\y,0-\z) -- (0.15+\y,-0.7-\z) -- (0.45+\y,-0.7-\z) -- (0.45+\y,0-\z);
        
         \fill[olive,opacity=0.8,rounded corners=3] (0,-.2-\z) -- (0.75,-.2-\z) -- (0.75,0.2-\z) -- (-0.75,0.2-\z) -- (-0.75,-.2-\z) -- (0,-.2-\z);
         \foreach \x in {-0.45,-0.15,...,0.75}
         \fill[black] (\x,-\z) circle (0.075);
         \fill[olive,opacity=0.8,rounded corners=3] (0+\y,-.2-\z) -- (0.75+\y,-.2-\z) -- (0.75+\y,0.2-\z) -- (-0.75+\y,0.2-\z) -- (-0.75+\y,-.2-\z) -- (0+\y,-.2-\z); 
         \foreach \x in {-0.45,-0.15,...,0.75}
         \fill[black] (\x+\y,-\z) circle (0.075);
        \end{tikzpicture}
    \end{array}.
\ee
This clearly must be a tensor product of two $|0\rangle$ states as directly follows from the definitions and from the calculations in the previous example.

\item Let us build an example of a nonmaximally entangled state. First, let us try the state
\be
\label{whatisN}
   |\psi_3\rangle \ = \ \begin{array}{c}
        \begin{tikzpicture}
            \newcommand{\y}{2}
         \newcommand{\z}{0.37}
         \draw[draw=white,double=black,line width=1,rounded corners=2] (-0.45,0-\z) -- (-0.45,-0.7-\z) -- (-0.15,-0.7-\z) -- (-0.15,0-\z);
         \draw[draw=white,double=black,line width=1,rounded corners=2] (0.15,-\z) -- (0.15,-0.7-\z) -- (-0.15+\y,-0.7-\z) -- (-0.15+\y,0-\z);
         \draw[draw=white,double=black,line width=1,rounded corners=2] (0.45,0-\z) -- (0.45,-0.5-\z) -- (-0.45+\y,-0.5-\z) -- (-0.45+\y,0-\z);
         \draw[draw=white,double=black,line width=1,rounded corners=2] (0.15+\y,0-\z) -- (0.15+\y,-0.7-\z) -- (0.45+\y,-0.7-\z) -- (0.45+\y,0-\z);
        
         \fill[olive,opacity=0.8,rounded corners=3] (0,-.2-\z) -- (0.75,-.2-\z) -- (0.75,0.2-\z) -- (-0.75,0.2-\z) -- (-0.75,-.2-\z) -- (0,-.2-\z);
         \foreach \x in {-0.45,-0.15,...,0.75}
         \fill[black] (\x,-\z) circle (0.075);
         \fill[olive,opacity=0.8,rounded corners=3] (0+\y,-.2-\z) -- (0.75+\y,-.2-\z) -- (0.75+\y,0.2-\z) -- (-0.75+\y,0.2-\z) -- (-0.75+\y,-.2-\z) -- (0+\y,-.2-\z); 
         \foreach \x in {-0.45,-0.15,...,0.75}
         \fill[black] (\x+\y,-\z) circle (0.075);
        \end{tikzpicture}
    \end{array}\,.
\ee
Using the method demonstrated by the first example it is not hard to see that the only nonzero overlap is with state $\langle00|$:
\be
    \langle 00|\psi_3\rangle \ = \ \frac{1}{d^2}
    \begin{array}{c}
        \begin{tikzpicture}
        \newcommand{\y}{2}
        \newcommand{\z}{0.37}
         \draw[draw=white,double=black,line width=1,rounded corners=2] (-0.45,0-\z) -- (-0.45,-0.7-\z) -- (-0.15,-0.7-\z) -- (-0.15,0-\z);
         \draw[draw=white,double=black,line width=1,rounded corners=2] (0.15,-\z) -- (0.15,-0.7-\z) -- (-0.15+\y,-0.7-\z) -- (-0.15+\y,0-\z);
         \draw[draw=white,double=black,line width=1,rounded corners=2] (0.45,0-\z) -- (0.45,-0.5-\z) -- (-0.45+\y,-0.5-\z) -- (-0.45+\y,0-\z);
         \draw[draw=white,double=black,line width=1,rounded corners=2] (0.15+\y,0-\z) -- (0.15+\y,-0.7-\z) -- (0.45+\y,-0.7-\z) -- (0.45+\y,0-\z);
        
         \fill[olive,opacity=0.4,rounded corners=3] (0,-.2-\z) -- (0.75,-.2-\z) -- (0.75,0.2-\z) -- (-0.75,0.2-\z) -- (-0.75,-.2-\z) -- (0,-.2-\z);
         \foreach \x in {-0.45,-0.15,...,0.75}
         \fill[black] (\x,-\z) circle (0.075);
         \fill[olive,opacity=0.4,rounded corners=3] (0+\y,-.2-\z) -- (0.75+\y,-.2-\z) -- (0.75+\y,0.2-\z) -- (-0.75+\y,0.2-\z) -- (-0.75+\y,-.2-\z) -- (0+\y,-.2-\z); 
         \foreach \x in {-0.45,-0.15,...,0.75}
         \fill[black] (\x+\y,-\z) circle (0.075);
         
         \draw[draw=white,double=black,line width=1,rounded corners=2] (-0.45,0-\z) -- (-0.45,0.7-\z) -- (-0.15,0.7-\z) -- (-0.15,0-\z);
         \draw[draw=white,double=black,line width=1,rounded corners=2] (0.15,-\z) -- (0.15,0.7-\z) -- (0.45,0.7-\z) -- (0.45,0-\z);
         \draw[draw=white,double=black,line width=1,rounded corners=2] (-0.45+\y,0-\z) -- (-0.45+\y,0.7-\z) -- (-0.15+\y,0.7-\z) -- (-0.15+\y,0-\z);
         \draw[draw=white,double=black,line width=1,rounded corners=2] (0.15+\y,-\z) -- (0.15+\y,0.7-\z) -- (0.45+\y,0.7-\z) -- (0.45+\y,0-\z);
        \end{tikzpicture}
    \end{array}
    \ = \ \frac{1}{d^2}\,d^3 \ = \ d\,.
\ee
In other words, $|\psi_3\rangle=d|00\rangle$ is a separable state.

This example illustrates a special property of Chern-Simons on two-spheres used in this work: if the space between two parties can be cut by an $S^2$ pierced by only two Wilson lines then the parties are separable. In this case we can show this using the following diagram:
\be
   \begin{array}{c}
        \begin{tikzpicture}
            \newcommand{\y}{2}
         \newcommand{\z}{0.37}
         \draw[draw=white,double=black,line width=1,rounded corners=2] (-0.45,0-\z) -- (-0.45,-0.7-\z) -- (-0.15,-0.7-\z) -- (-0.15,0-\z);
         \draw[draw=white,double=black,line width=1,rounded corners=2] (0.15,-\z) -- (0.15,-0.7-\z) -- (-0.15+\y,-0.7-\z) -- (-0.15+\y,0-\z);
         \draw[draw=white,double=black,line width=1,rounded corners=2] (0.45,0-\z) -- (0.45,-0.5-\z) -- (-0.45+\y,-0.5-\z) -- (-0.45+\y,0-\z);
         \draw[draw=white,double=black,line width=1,rounded corners=2] (0.15+\y,0-\z) -- (0.15+\y,-0.7-\z) -- (0.45+\y,-0.7-\z) -- (0.45+\y,0-\z);
        
         \fill[olive,opacity=0.8,rounded corners=3] (0,-.2-\z) -- (0.75,-.2-\z) -- (0.75,0.2-\z) -- (-0.75,0.2-\z) -- (-0.75,-.2-\z) -- (0,-.2-\z);
         \foreach \x in {-0.45,-0.15,...,0.75}
         \fill[black] (\x,-\z) circle (0.075);
         \fill[olive,opacity=0.8,rounded corners=3] (0+\y,-.2-\z) -- (0.75+\y,-.2-\z) -- (0.75+\y,0.2-\z) -- (-0.75+\y,0.2-\z) -- (-0.75+\y,-.2-\z) -- (0+\y,-.2-\z); 
         \foreach \x in {-0.45,-0.15,...,0.75}
         \fill[black] (\x+\y,-\z) circle (0.075);

          \fill[olive,opacity=0.4,rounded corners=3] (0,0.4-\z) -- (0.95,0.4-\z) -- (0.95,-0.9-\z) -- (-0.95,-0.9-\z) -- (-0.95,.4-\z) -- (0,0.4-\z);
          \fill[olive,opacity=0.8,rounded corners=3] (0.95,-0.7-\z) -- (0.95,-0.85-\z) -- (0.65,-0.85-\z) -- (0.65,-0.35-\z) -- (0.95,-.35-\z) -- (0.95,-0.7-\z);
          \fill[black] (0.8,-0.5-\z) circle (0.075);
          \fill[black] (0.8,-0.7-\z) circle (0.075);
          \draw[draw=white,double=black,line width=1,rounded corners=2] (0.8,-0.5-\z) -- (1,-0.5-\z);
          \draw[draw=white,double=black,line width=1,rounded corners=2] (0.8,-0.7-\z) -- (1,-0.7-\z);
        \end{tikzpicture}
    \end{array}\ = \ 
    \frac{1}{d}
       \begin{array}{c}
        \begin{tikzpicture}
            \newcommand{\y}{2}
         \newcommand{\z}{0.37}
         \draw[draw=white,double=black,line width=1,rounded corners=2] (-0.45,0-\z) -- (-0.45,-0.7-\z) -- (-0.15,-0.7-\z) -- (-0.15,0-\z);
         \draw[draw=white,double=black,line width=1,rounded corners=2] (0.15,0-\z) -- (0.15,-0.7-\z) -- (0.45,-0.7-\z) -- (0.45,0-\z);
         \draw[draw=white,double=black,line width=1,rounded corners=2] (-0.45+\y,0-\z) -- (-0.45+\y,-0.7-\z) -- (-0.15+\y,-0.7-\z) -- (-0.15+\y,0-\z);
         \draw[draw=white,double=black,line width=1,rounded corners=2] (0.15+\y,0-\z) -- (0.15+\y,-0.7-\z) -- (0.45+\y,-0.7-\z) -- (0.45+\y,0-\z);
        
         \fill[olive,opacity=0.8,rounded corners=3] (0,-.2-\z) -- (0.75,-.2-\z) -- (0.75,0.2-\z) -- (-0.75,0.2-\z) -- (-0.75,-.2-\z) -- (0,-.2-\z);
         \foreach \x in {-0.45,-0.15,...,0.75}
         \fill[black] (\x,-\z) circle (0.075);
         \fill[olive,opacity=0.8,rounded corners=3] (0+\y,-.2-\z) -- (0.75+\y,-.2-\z) -- (0.75+\y,0.2-\z) -- (-0.75+\y,0.2-\z) -- (-0.75+\y,-.2-\z) -- (0+\y,-.2-\z); 
         \foreach \x in {-0.45,-0.15,...,0.75}
         \fill[black] (\x+\y,-\z) circle (0.075);
        \end{tikzpicture}
    \end{array}.
\ee
The left sphere can be inserted inside a larger 2-sphere, which is only cut by two Wilson lines. The dimension of the Hilbert space associated with a sphere with two punctures is one (if the punctures are labeled by conjugate representations, e.g. two spin one-half irreps, and zero otherwise~\cite{Witten:1988hf}), so one cut the two lines as the above diagram shows. Equivalently, one think of this cutting as inserting a completeness relation in a one-dimensional Hilbert space:
\be
\mathbb{I}_1 \ = \ \begin{array}{c}
        \begin{tikzpicture}
         \newcommand{\z}{0}
         \newcommand{\x}{1.5}
         \fill[olive,opacity=0.8,rounded corners=3] (0.95,-0.7-\z) -- (0.95,-0.85-\z) -- (0.65,-0.85-\z) -- (0.65,-0.35-\z) -- (0.95,-.35-\z) -- (0.95,-0.7-\z);
          \fill[black] (0.8,-0.5-\z) circle (0.075);
          \fill[black] (0.8,-0.7-\z) circle (0.075);

          \fill[olive,opacity=0.8,rounded corners=3] (0.95+\x,-0.7-\z) -- (0.95+\x,-0.85-\z) -- (0.65+\x,-0.85-\z) -- (0.65+\x,-0.35-\z) -- (0.95+\x,-.35-\z) -- (0.95+\x,-0.7-\z);
          \fill[black] (0.8+\x,-0.5-\z) circle (0.075);
          \fill[black] (0.8+\x,-0.7-\z) circle (0.075);
          \draw (0.8,-0.5-\z) -- (0.8+\x,-0.5-\z);
          \draw (0.8,-0.7-\z) -- (0.8+\x,-0.7-\z);
        \end{tikzpicture}
    \end{array}
    \ = \ 
    \begin{array}{c}
        \begin{tikzpicture}
         \newcommand{\z}{0}
         \newcommand{\x}{1}
         \fill[olive,opacity=0.8,rounded corners=3] (0.95,-0.7-\z) -- (0.95,-0.85-\z) -- (0.65,-0.85-\z) -- (0.65,-0.35-\z) -- (0.95,-.35-\z) -- (0.95,-0.7-\z);
          \fill[black] (0.8,-0.5-\z) circle (0.075);
          \fill[black] (0.8,-0.7-\z) circle (0.075);
           \fill[olive,opacity=0.4,rounded corners=3] (0.95+\x,-0.7-\z) -- (0.95+\x,-0.85-\z) -- (0.65+\x,-0.85-\z) -- (0.65+\x,-0.35-\z) -- (0.95+\x,-.35-\z) -- (0.95+\x,-0.7-\z);
          \fill[black] (0.8+\x,-0.5-\z) circle (0.075);
          \fill[black] (0.8+\x,-0.7-\z) circle (0.075);
          \draw (0.8,-0.5-\z) -- (0.8+\x,-0.5-\z);
          \draw (0.8,-0.7-\z) -- (0.8+\x,-0.7-\z);
          \draw[rounded corners=3] (0.8+\x,-0.7-\z) -- (1.2+\x,-0.7-\z) -- (1.2+\x,-0.5-\z) -- (0.8+\x,-0.5-\z);

        \renewcommand{\x}{2}
          \fill[olive,opacity=0.4,rounded corners=3] (0.95+\x,-0.7-\z) -- (0.95+\x,-0.85-\z) -- (0.65+\x,-0.85-\z) -- (0.65+\x,-0.35-\z) -- (0.95+\x,-.35-\z) -- (0.95+\x,-0.7-\z);
          \fill[black] (0.8+\x,-0.5-\z) circle (0.075);
          \fill[black] (0.8+\x,-0.7-\z) circle (0.075);
          \draw (1.8+\x,-0.5-\z) -- (0.8+\x,-0.5-\z);
          \draw (1.8+\x,-0.7-\z) -- (0.8+\x,-0.7-\z);
          \draw[rounded corners=3] (0.8+\x,-0.7-\z) -- (0.4+\x,-0.7-\z) -- (0.4+\x,-0.5-\z) -- (0.8+\x,-0.5-\z);

        \renewcommand{\x}{3}
          \fill[olive,opacity=0.8,rounded corners=3] (0.95+\x,-0.7-\z) -- (0.95+\x,-0.85-\z) -- (0.65+\x,-0.85-\z) -- (0.65+\x,-0.35-\z) -- (0.95+\x,-.35-\z) -- (0.95+\x,-0.7-\z);
          \fill[black] (0.8+\x,-0.5-\z) circle (0.075);
          \fill[black] (0.8+\x,-0.7-\z) circle (0.075);
        \end{tikzpicture} 
    \end{array} \ = \ |0_1\rangle\langle 0_1|\,. 
\ee
In order to build a nonmaximally entangled and nonseparable state a more complex topology is required.

\item Let us consider a state~\cite{Dwivedi:2019bzh,Melnikov:2023wwc}:
\be
    |\psi_4\rangle \ = \ %
\ee
To compute the basis expansion of this state we will need to use skein relation~(\ref{skein}):
\be
    
    \ = \
    A 
    \begin{array}{c}
    \begin{tikzpicture}
        \newcommand{\x}{0.6}
        \newcommand{\y}{2}
        \newcommand{\z}{0.37}
        
        \draw[draw=white,double=black,line width=1,rounded corners=2] (-0.45,0-\z) -- (-0.45,-0.68-\z-.45) -- (1.2,-1.5) (1.55,-0.7-\z-.45) -- (0.45+\y,-0.7-\z-.45) -- (0.45+\y,0-\z);
        
        \draw[draw=white,double=black,line width=1,rounded corners=2] (-0.15,0-\z) -- (-0.15,-0.6-\z) -- (.45,-.6-\z) -- (.8,-.8-\z) arc (0:-90:0.175cm) (1.55,-.6-\z) -- (0.15+\y,-0.6-\z) -- (0.15+\y,0-\z);

        \draw[draw=white,double=black,line width=1,rounded corners=2] (0.15,-\z) -- (0.15,-0.5-\z) -- (-0.15+\y,-0.5-\z) -- (-0.15+\y,0-\z);
        \draw[draw=white,double=black,line width=1,rounded corners=2] (0.45,0-\z) -- (0.45,-0.4-\z) -- (-0.45+\y,-0.4-\z) -- (-0.45+\y,0-\z);
        
        \draw[draw=white,double=black,line width=1,rounded corners=2] (1.55,-.6-\z) arc (90:270:0.275cm);
        \draw[draw=white,double=black,line width=1,rounded corners=2] (1.2,-1.5) arc (-90:30:0.275cm);
        \draw[draw=white,double=black,line width=1,rounded corners=2] (1.2,-.6-\z) arc (90:60:0.275cm);
        
        \draw[draw=white,double=black,line width=1,rounded corners=2] (1.2,-.6-\z) -- (.8,-.6-\z) -- (.45,-.8-\z) arc (180:270:0.175cm);

        \fill[olive,opacity=0.8,rounded corners=3] (0,-.2-\z) -- (0.75,-.2-\z) -- (0.75,0.2-\z) -- (-0.75,0.2-\z) -- (-0.75,-.2-\z) -- (0,-.2-\z);
        \foreach \x in {-0.45,-0.15,...,0.75}
        \fill[black] (\x,-\z) circle (0.075);
        \fill[olive,opacity=0.8,rounded corners=3] (0+\y,-.2-\z) -- (0.75+\y,-.2-\z) -- (0.75+\y,0.2-\z) -- (-0.75+\y,0.2-\z) -- (-0.75+\y,-.2-\z) -- (0+\y,-.2-\z); 
        \foreach \x in {-0.45,-0.15,...,0.75}
        \fill[black] (\x+\y,-\z) circle (0.075);
    \end{tikzpicture}
    \end{array}
    \ + A^{-1}
    \begin{array}{c}
        \begin{tikzpicture}
            \newcommand{\y}{2}
         \newcommand{\z}{0.37}
         
         \draw[draw=white,double=black,line width=1,rounded corners=2] (-0.45,0-\z) -- (-0.45,-1.5) -- (-1.50+\y,-1.5)--(-1.15+\y,-0.6-\z)--(-0.8+\y,-0.6-\z);
         \draw[draw=white,double=black,line width=1,rounded corners=2] (-0.15,0-\z) -- (-0.15,-0.6-\z) -- (-1.50+\y,-0.6-\z)--(-1.15+\y,-1.5)--(-0.8+\y,-1.5);

         \draw[draw=white,double=black,line width=1,rounded corners=2] (-0.5+\y,-1.5) arc (-90:-210:0.265cm);
         \draw[draw=white,double=black,line width=1,rounded corners=2](-0.8+\y,-1.5) arc (-90:90:0.265cm);
         \draw[draw=white,double=black,line width=1,rounded corners=2] (0.45+\y,0-\z) -- (0.45+\y,-1.5) -- (-0.5+\y,-1.5);
         \draw[draw=white,double=black,line width=1,rounded corners=2] (0.15+\y,0-\z) -- (0.15+\y,-0.6-\z) -- (-0.5+\y,-0.6-\z);
         \draw[draw=white,double=black,line width=1,rounded corners=2] (-0.5+\y,-0.6-\z) arc (90:210:0.265cm);

         \draw[draw=white,double=black,line width=1,rounded corners=2] (0.15,-\z) -- (0.15,-0.5-\z) -- (-0.15+\y,-0.5-\z) -- (-0.15+\y,0-\z);
         \draw[draw=white,double=black,line width=1,rounded corners=2] (0.45,0-\z) -- (0.45,-0.4-\z) -- (-0.45+\y,-0.4-\z) -- (-0.45+\y,0-\z);
        
         \fill[olive,opacity=0.8,rounded corners=3] (0,-.2-\z) -- (0.75,-.2-\z) -- (0.75,0.2-\z) -- (-0.75,0.2-\z) -- (-0.75,-.2-\z) -- (0,-.2-\z);
         \foreach \x in {-0.45,-0.15,...,0.75}
         \fill[black] (\x,-\z) circle (0.075);
         \fill[olive,opacity=0.8,rounded corners=3] (0+\y,-.2-\z) -- (0.75+\y,-.2-\z) -- (0.75+\y,0.2-\z) -- (-0.75+\y,0.2-\z) -- (-0.75+\y,-.2-\z) -- (0+\y,-.2-\z); 
         \foreach \x in {-0.45,-0.15,...,0.75}
         \fill[black] (\x+\y,-\z) circle (0.075);
        \end{tikzpicture}
    \end{array} \,,   
\ee
here applied to the bottom left crossing. By successive application of the skein relation on the other crossings yields
\be
    \begin{aligned}
        
    
    \\ 
    & + \ A^{-2}
    \begin{array}{c}
    \begin{tikzpicture}
        \newcommand{\x}{0.6}
        \newcommand{\y}{2}
        \newcommand{\z}{0.37}
        \draw[draw=white,double=black,line width=1,rounded corners=2,xscale=-1] (-0.45,0-\z) -- (-0.45,-0.68-\z-.45) -- (1.2,-1.5)(1.2,-.6-\z) -- (1.55,-0.7-\z-.45) -- (0.45+\y,-0.7-\z-.45) -- (0.45+\y,0-\z);
        
        \draw[draw=white,double=black,line width=1,rounded corners=2,xscale=-1] (-0.15,0-\z) -- (-0.15,-0.6-\z) -- (.45,-.6-\z) -- (.8,-.8-\z) arc (0:-90:0.175cm) (1.2,-1.5) -- (1.55,-.6-\z) -- (0.15+\y,-0.6-\z) -- (0.15+\y,0-\z);
        
        \draw[draw=white,double=black,line width=1,rounded corners=2,xscale=-1] (0.15,-\z) -- (0.15,-0.5-\z) -- (-0.15+\y,-0.5-\z) -- (-0.15+\y,0-\z);
        \draw[draw=white,double=black,line width=1,rounded corners=2,xscale=-1] (0.45,0-\z) -- (0.45,-0.4-\z) -- (-0.45+\y,-0.4-\z) -- (-0.45+\y,0-\z);

        \draw[draw=white,double=black,line width=1,rounded corners=2,xscale=-1] (1.2,-.6-\z) -- (.8,-.6-\z) -- (.45,-.8-\z) arc (180:270:0.175cm);

        \fill[olive,opacity=0.8,rounded corners=3] (0,-.2-\z) -- (0.75,-.2-\z) -- (0.75,0.2-\z) -- (-0.75,0.2-\z) -- (-0.75,-.2-\z) -- (0,-.2-\z);
        \foreach \x in {-0.45,-0.15,...,0.75}
        \fill[black] (\x,-\z) circle (0.075);

        \fill[olive,opacity=0.8,rounded corners=3] (-\y,-.2-\z) -- (-0.75-\y,-.2-\z) -- (-.75-\y,0.2-\z) -- (+.75-\y,0.2-\z) -- (+0.75-\y,-.2-\z) -- (-\y,-.2-\z); 
        \foreach \x in {-0.45,-0.15,...,0.75}
        \fill[black] (-\x-\y,-\z) circle (0.075);
        
    \end{tikzpicture}
    \end{array}
    +\ 
    \begin{array}{c}
    \begin{tikzpicture}
        \newcommand{\x}{0.6}
        \newcommand{\y}{2}
        \newcommand{\z}{0.37}
        \draw[draw=white,double=black,line width=1,rounded corners=2,xscale=-1] (-0.45,0-\z) -- (-0.45,-0.68-\z-.45) -- (0.45,-0.68-\z-.45)--(.75,-.6-\z) --(1.2,-.6-\z) -- (1.55,-0.7-\z-.45) -- (0.45+\y,-0.7-\z-.45) -- (0.45+\y,0-\z);
        
        \draw[draw=white,double=black,line width=1,rounded corners=2,xscale=-1] (-0.15,0-\z) -- (-0.15,-0.6-\z) -- (.45,-.6-\z) -- (.75,-1.5) -- (1.2,-1.5) -- (1.55,-.6-\z) -- (0.15+\y,-0.6-\z) -- (0.15+\y,0-\z);
        
        \draw[draw=white,double=black,line width=1,rounded corners=2,xscale=-1] (0.15,-\z) -- (0.15,-0.5-\z) -- (-0.15+\y,-0.5-\z) -- (-0.15+\y,0-\z);
        \draw[draw=white,double=black,line width=1,rounded corners=2,xscale=-1] (0.45,0-\z) -- (0.45,-0.4-\z) -- (-0.45+\y,-0.4-\z) -- (-0.45+\y,0-\z);

        \fill[olive,opacity=0.8,rounded corners=3] (0,-.2-\z) -- (0.75,-.2-\z) -- (0.75,0.2-\z) -- (-0.75,0.2-\z) -- (-0.75,-.2-\z) -- (0,-.2-\z);
        \foreach \x in {-0.45,-0.15,...,0.75}
        \fill[black] (\x,-\z) circle (0.075);

        \fill[olive,opacity=0.8,rounded corners=3] (-\y,-.2-\z) -- (-0.75-\y,-.2-\z) -- (-.75-\y,0.2-\z) -- (+.75-\y,0.2-\z) -- (+0.75-\y,-.2-\z) -- (-\y,-.2-\z); 
        \foreach \x in {-0.45,-0.15,...,0.75}
        \fill[black] (-\x-\y,-\z) circle (0.075);
        
    \end{tikzpicture}
\end{array}\\
    & =\ \left(A^{2} - A^{-2}\right)^2
    \begin{array}{c}
    \begin{tikzpicture}
        \newcommand{\y}{2}
         \newcommand{\z}{0.37}
         \draw[draw=white,double=black,line width=1,rounded corners=2] (-0.45,0-\z) -- (-0.45,-0.7-\z) -- (0.45+\y,-0.7-\z) -- (0.45+\y,0-\z);
         \draw[draw=white,double=black,line width=1,rounded corners=2] (-0.15,0-\z) -- (-0.15,-0.6-\z) -- (0.15+\y,-0.6-\z) -- (0.15+\y,0-\z);
         \draw[draw=white,double=black,line width=1,rounded corners=2] (0.15,-\z) -- (0.15,-0.5-\z) -- (-0.15+\y,-0.5-\z) -- (-0.15+\y,0-\z);
         \draw[draw=white,double=black,line width=1,rounded corners=2] (0.45,0-\z) -- (0.45,-0.4-\z) -- (-0.45+\y,-0.4-\z) -- (-0.45+\y,0-\z);
        
         \fill[olive,opacity=0.8,rounded corners=3] (0,-.2-\z) -- (0.75,-.2-\z) -- (0.75,0.2-\z) -- (-0.75,0.2-\z) -- (-0.75,-.2-\z) -- (0,-.2-\z);
         \foreach \x in {-0.45,-0.15,...,0.75}
         \fill[black] (\x,-\z) circle (0.075);
         \fill[olive,opacity=0.8,rounded corners=3] (0+\y,-.2-\z) -- (0.75+\y,-.2-\z) -- (0.75+\y,0.2-\z) -- (-0.75+\y,0.2-\z) -- (-0.75+\y,-.2-\z) -- (0+\y,-.2-\z); 
         \foreach \x in {-0.45,-0.15,...,0.75}
         \fill[black] (\x+\y,-\z) circle (0.075);
    \end{tikzpicture}
    \end{array}\\
    \ & + \
    \left(A^6 + A^{-6}\right)
    \begin{array}{c}
        \begin{tikzpicture}
            \newcommand{\y}{2}
         \newcommand{\z}{0.37}
         \draw[draw=white,double=black,line width=1,rounded corners=2] (-0.45,0-\z) -- (-0.45,-0.7-\z) -- (-0.15,-0.7-\z) -- (-0.15,0-\z);
         \draw[draw=white,double=black,line width=1,rounded corners=2] (0.15,-\z) -- (0.15,-0.7-\z) -- (-0.15+\y,-0.7-\z) -- (-0.15+\y,0-\z);
         \draw[draw=white,double=black,line width=1,rounded corners=2] (0.45,0-\z) -- (0.45,-0.5-\z) -- (-0.45+\y,-0.5-\z) -- (-0.45+\y,0-\z);
         \draw[draw=white,double=black,line width=1,rounded corners=2] (0.15+\y,0-\z) -- (0.15+\y,-0.7-\z) -- (0.45+\y,-0.7-\z) -- (0.45+\y,0-\z);
        
         \fill[olive,opacity=0.8,rounded corners=3] (0,-.2-\z) -- (0.75,-.2-\z) -- (0.75,0.2-\z) -- (-0.75,0.2-\z) -- (-0.75,-.2-\z) -- (0,-.2-\z);
         \foreach \x in {-0.45,-0.15,...,0.75}
         \fill[black] (\x,-\z) circle (0.075);
         \fill[olive,opacity=0.8,rounded corners=3] (0+\y,-.2-\z) -- (0.75+\y,-.2-\z) -- (0.75+\y,0.2-\z) -- (-0.75+\y,0.2-\z) -- (-0.75+\y,-.2-\z) -- (0+\y,-.2-\z); 
         \foreach \x in {-0.45,-0.15,...,0.75}
         \fill[black] (\x+\y,-\z) circle (0.075);
        \end{tikzpicture}
    \end{array}

    \end{aligned}
\label{ex4-2}
\ee

Here the loop in the first term of (\ref{ex4}) has been replaced by $(-A^{3})$, and in (\ref{ex4-2}) -- by $(-A^{-3})$. Thus,
\be
\begin{aligned}
     
    & =\ 
    \ \left(A^{2} - A^{-2}\right)^2
    \begin{array}{c}
    \begin{tikzpicture}
        \newcommand{\y}{2}
         \newcommand{\z}{0.37}
         \draw[draw=white,double=black,line width=1,rounded corners=2] (-0.45,0-\z) -- (-0.45,-0.7-\z) -- (0.45+\y,-0.7-\z) -- (0.45+\y,0-\z);
         \draw[draw=white,double=black,line width=1,rounded corners=2] (-0.15,0-\z) -- (-0.15,-0.6-\z) -- (0.15+\y,-0.6-\z) -- (0.15+\y,0-\z);
         \draw[draw=white,double=black,line width=1,rounded corners=2] (0.15,-\z) -- (0.15,-0.5-\z) -- (-0.15+\y,-0.5-\z) -- (-0.15+\y,0-\z);
         \draw[draw=white,double=black,line width=1,rounded corners=2] (0.45,0-\z) -- (0.45,-0.4-\z) -- (-0.45+\y,-0.4-\z) -- (-0.45+\y,0-\z);
        
         \fill[olive,opacity=0.8,rounded corners=3] (0,-.2-\z) -- (0.75,-.2-\z) -- (0.75,0.2-\z) -- (-0.75,0.2-\z) -- (-0.75,-.2-\z) -- (0,-.2-\z);
         \foreach \x in {-0.45,-0.15,...,0.75}
         \fill[black] (\x,-\z) circle (0.075);
         \fill[olive,opacity=0.8,rounded corners=3] (0+\y,-.2-\z) -- (0.75+\y,-.2-\z) -- (0.75+\y,0.2-\z) -- (-0.75+\y,0.2-\z) -- (-0.75+\y,-.2-\z) -- (0+\y,-.2-\z); 
         \foreach \x in {-0.45,-0.15,...,0.75}
         \fill[black] (\x+\y,-\z) circle (0.075);
    \end{tikzpicture}
    \end{array}\\
    & +\
    \frac{\left(A^6+A^{-6}\right)}{d}
    \begin{array}{c}
    \begin{tikzpicture}
        \newcommand{\y}{2}
         \newcommand{\z}{0.37}
         \draw[draw=white,double=black,line width=1,rounded corners=2] (-0.45,0-\z) -- (-0.45,-0.7-\z) -- (-0.15,-0.7-\z) -- (-0.15,0-\z);
         \draw[draw=white,double=black,line width=1,rounded corners=2] (0.15,0-\z) -- (0.15,-0.7-\z) -- (0.45,-0.7-\z) -- (0.45,0-\z);
         \draw[draw=white,double=black,line width=1,rounded corners=2] (-0.45+\y,0-\z) -- (-0.45+\y,-0.7-\z) -- (-0.15+\y,-0.7-\z) -- (-0.15+\y,0-\z);
         \draw[draw=white,double=black,line width=1,rounded corners=2] (0.15+\y,0-\z) -- (0.15+\y,-0.7-\z) -- (0.45+\y,-0.7-\z) -- (0.45+\y,0-\z);
        
         \fill[olive,opacity=0.8,rounded corners=3] (0,-.2-\z) -- (0.75,-.2-\z) -- (0.75,0.2-\z) -- (-0.75,0.2-\z) -- (-0.75,-.2-\z) -- (0,-.2-\z);
         \foreach \x in {-0.45,-0.15,...,0.75}
         \fill[black] (\x,-\z) circle (0.075);
         \fill[olive,opacity=0.8,rounded corners=3] (0+\y,-.2-\z) -- (0.75+\y,-.2-\z) -- (0.75+\y,0.2-\z) -- (-0.75+\y,0.2-\z) -- (-0.75+\y,-.2-\z) -- (0+\y,-.2-\z); 
         \foreach \x in {-0.45,-0.15,...,0.75}
         \fill[black] (\x+\y,-\z) circle (0.075);
    \end{tikzpicture}
    \end{array} 
\end{aligned}
\ee
where the result from the previous example was applied in the last term. Finally, it is straightforward to obtain the expansion in the computational basis~(\ref{ONbasis}):
\be     
|\psi_4\rangle = \left(A^4+A^{-4}\right)^2|00\rangle + (A^2-A^{-2})^2|11\rangle\,.
\ee
Thus the result is indeed a nonmaximaly entangled state.

    \section{Topological encoding in the computational basis}
\label{sec:coefficients}

In this appendix we collect some additional details about state~(\ref{encodingchip}) used for information recovery in section~\ref{sec:errcorr1}. As was explained, the encoding of a logical qubit can be realized by teleporting it to the central qubit of a 5-qubit chip as shown in the diagram below:
\be
\begin{array}{c}
     \begin{tikzpicture}
         \fill[gray,opacity=0.5,rounded corners=5] (0,-2) -- (2,-2) -- (2,2) -- (-2,2) -- (-2,-2) -- (0,-2);
         \fill[olive,rounded corners=3] (0,-2.3) -- (1,-2.3) -- (1,-1.7) -- (-1,-1.7) -- (-1,-2.3) -- (0,-2.3);
         \fill[olive,rounded corners=3] (0,2.3) -- (1,2.3) -- (1,1.7) -- (-1,1.7) -- (-1,2.3) -- (0,2.3);
         \fill[olive,rounded corners=3] (2.3,0) -- (2.3,1) -- (1.7,1) -- (1.7,-1) -- (2.3,-1) -- (2.3,0);
         \fill[olive,rounded corners=3] (-2.3,0) -- (-2.3,1) -- (-1.7,1) -- (-1.7,-1) -- (-2.3,-1) -- (-2.3,0);
         \fill[olive,rounded corners=3] (0,-.3) -- (1,-.3) -- (1,0.3) -- (-1,0.3) -- (-1,-.3) -- (0,-.3);
         \draw[line width=1.5,rounded corners=3] (-2,0.2) -- (-1.4,0.2) -- (-0.8,0.8) -- (-0.3,0.8);
         \draw[line width=1.5,rounded corners=3] (-0.1,0.8) -- (0.8,0.8) -- (1.4,0.2) -- (2,0.2);
         \draw[line width=1.5,rounded corners=3] (0.2,2) -- (0.2,1.2) -- (0.5,0.9);
         \draw[line width=1.5,rounded corners=3]  (0.7,0.7) -- (1.1,0.3) -- (1.1,-0.4) -- (1.0,-0.5);
         \draw[line width=1.5,rounded corners=3]  (0.8,-0.7) -- (0.2,-1.4) -- (0.2,-2);
         \draw[line width=1.5,rounded corners=3] (-0.6,-2) -- (-0.6,-1.4) -- (-1.4,-0.6) -- (-2,-0.6);
         \draw[line width=1.5,rounded corners=3] (0.6,-2) -- (0.6,-1.4) -- (1.4,-0.6) -- (2,-0.6);
         \draw[line width=1.5,rounded corners=3] (-0.6,2) -- (-0.6,1.4) -- (-1.4,0.6) -- (-2,0.6);
         \draw[line width=1.5,rounded corners=3] (0.6,2) -- (0.6,1.4) -- (1.4,0.6) -- (2,0.6);
         \draw[line width=1.5,rounded corners=3] (-2,-0.2) -- (-1.4,-0.2) -- (-1,-0.6) -- (-0.7,-0.6);
         \draw[line width=1.5,rounded corners=3] (-0.5,-0.6) -- (-0.2,-0.6) -- (-0.2,0);
         \draw[line width=1.5,rounded corners=3] (2,-0.2) -- (1.4,-0.2) -- (1.2,-0.6) -- (0.2,-0.6) -- (0.2,0);
         \draw[line width=1.5,rounded corners=3] (-0.2,-2) -- (-0.2,-1.4) -- (-0.6,-1) -- (-0.6,0);
         \draw[line width=1.5,rounded corners=3] (0.6,0) -- (0.6,0.6) -- (-0.2,0.6) -- (-0.2,2);
         \foreach \x in {-0.6,-0.2,...,1}
           \fill[black] (\x,0) circle (0.075);
           \foreach \x in {-0.6,-0.2,...,1}
           \fill[black] (\x,-2) circle (0.075);
           \foreach \x in {-0.6,-0.2,...,1}
           \fill[black] (\x,2) circle (0.075);
           \foreach \x in {-0.6,-0.2,...,1}
           \fill[black] (-2,\x) circle (0.075);
           \foreach \x in {-0.6,-0.2,...,1}
           \fill[black] (2,\x) circle (0.075);
            \newcommand{\y}{-4}
            \newcommand{\z}{-0.2}
            \draw[draw=white,double=black,line width=1,rounded corners=2] (0+\y,-0.55+\z) -- (0.7+\y,-0.55+\z) -- (0.7+\y,-0.65+\z) -- (-0.7+\y,-0.65+\z) -- (-0.7+\y,-0.55+\z) -- (0+\y,-0.55+\z);
       \draw[draw=white,double=black,line width=1,rounded corners=2] (0.6+\y,-0.6+\z) -- (0.6+\y,-0.7+\z) -- (-0.6+\y,-0.7+\z) -- (-0.6+\y,-0.5+\z) -- (0.45+\y,-0.5+\z);
         \draw[draw=white,double=black,line width=1,rounded corners=2] (-0.05+\y,-0.5+\z) -- (0.1+\y,-0.5+\z) -- (0.1+\y,-0.7+\z) -- (0.35+\y,-0.7+\z) -- (.35+\y,-0.5+\z) -- (0.6+\y,-0.5+\z) -- (0.6+\y,-0.6+\z);
         \draw[draw=white,double=black,line width=1,rounded corners=2] (-0.6+\y,0+\z) -- (-0.6+\y,-0.45+\z) -- (-0.45+\y,-0.45+\z) -- (-0.45+\y,-0.7+\z) -- (0+\y,-0.7+\z);
         \draw[draw=white,double=black,line width=1,rounded corners=2] (-0.2+\y,0+\z) -- (-0.2+\y,-0.5+\z) -- (-0.35+\y,-0.5+\z) -- (-0.35+\y,-0.7+\z) -- (-0.55+\y,-0.7+\z) -- (-0.55+\y,-0.6+\z);
         \draw[draw=white,double=black,line width=1,rounded corners=2] (-0.5+\y,-0.5+\z) -- (-0.7+\y,-0.5+\z) -- (-0.7+\y,-0.6+\z) -- (0.25+\y,-0.6+\z) -- (0.25+\y,-0.5+\z) -- (0.2+\y,-0.5+\z) -- (0.2+\y,0+\z);
         \draw[draw=white,double=black,line width=1,rounded corners=2] (0.6+\y,0+\z) -- (0.6+\y,-0.45+\z) -- (0.45+\y,-0.45+\z) -- (0.45+\y,-0.6+\z);
         \fill[olive,rounded corners=3] (-4,-.3) -- (-3,-.3) -- (-3,0.3) -- (-5,0.3) -- (-5,-.3) -- (-4,-.3);
           \foreach \x in {-4.6,-4.2,...,-3}
           \fill[black] (\x,0) circle (0.075);
           \draw[orange,dashed,line width=1.5,rounded corners=3] (-0.6+\y,0) -- (-0.6+\y,1.2) -- (-0.6,1.2) -- (-0.6,0);
           \draw[orange,dashed,line width=1.5,rounded corners=3] (-0.2+\y,0) -- (-0.2+\y,1.3) -- (-0.5,1.3) -- (-0.5,0) -- (-0.2,0);
           \draw[orange,dashed,line width=1.5,rounded corners=3] (0.2+\y,0) -- (0.2+\y,1.4) -- (-0.4,1.4) -- (-0.4,0.1) -- (0.2,0.1) -- (0.2,0);
           \draw[orange,dashed,line width=1.5,rounded corners=3] (0.6+\y,0) -- (0.6+\y,1.5) -- (-0.3,1.5) -- (-0.3,0.2) -- (0.4,0.2) -- (0.4,0) -- (0.6,0);
            \fill[orange,rounded corners=3] (-4,.6) -- (-3,.6) -- (-3,1) -- (-5,1) -- (-5,.6) -- (-4,.6);
     \end{tikzpicture} 
\end{array}
\quad\Longrightarrow\quad
\begin{array}{c}
     \begin{tikzpicture}
         \fill[gray,opacity=0.5,rounded corners=5] (0,-2) -- (2,-2) -- (2,2) -- (-2,2) -- (-2,-2) -- (0,-2);
         \fill[olive,rounded corners=3] (0,-2.3) -- (1,-2.3) -- (1,-1.7) -- (-1,-1.7) -- (-1,-2.3) -- (0,-2.3);
         \fill[olive,rounded corners=3] (0,2.3) -- (1,2.3) -- (1,1.7) -- (-1,1.7) -- (-1,2.3) -- (0,2.3);
         \fill[olive,rounded corners=3] (2.3,0) -- (2.3,1) -- (1.7,1) -- (1.7,-1) -- (2.3,-1) -- (2.3,0);
         \fill[olive,rounded corners=3] (-2.3,0) -- (-2.3,1) -- (-1.7,1) -- (-1.7,-1) -- (-2.3,-1) -- (-2.3,0);
         \fill[white,rounded corners=3] (0,-.3) -- (1,-.3) -- (1,0.3) -- (-1,0.3) -- (-1,-.3) -- (0,-.3);
         \draw[line width=1.5,rounded corners=3] (-2,0.2) -- (-1.4,0.2) -- (-0.8,0.8) -- (-0.3,0.8);
         \draw[line width=1.5,rounded corners=3] (-0.1,0.8) -- (0.8,0.8) -- (1.4,0.2) -- (2,0.2);
         \draw[line width=1.5,rounded corners=3] (0.2,2) -- (0.2,1.2) -- (0.5,0.9);
         \draw[line width=1.5,rounded corners=3]  (0.7,0.7) -- (1.1,0.3) -- (1.1,-0.4) -- (1.0,-0.5);
         \draw[line width=1.5,rounded corners=3]  (0.8,-0.7) -- (0.2,-1.4) -- (0.2,-2);
         \draw[line width=1.5,rounded corners=3] (-0.6,-2) -- (-0.6,-1.4) -- (-1.4,-0.6) -- (-2,-0.6);
         \draw[line width=1.5,rounded corners=3] (0.6,-2) -- (0.6,-1.4) -- (1.4,-0.6) -- (2,-0.6);
         \draw[line width=1.5,rounded corners=3] (-0.6,2) -- (-0.6,1.4) -- (-1.4,0.6) -- (-2,0.6);
         \draw[line width=1.5,rounded corners=3] (0.6,2) -- (0.6,1.4) -- (1.4,0.6) -- (2,0.6);
         \draw[line width=1.5,rounded corners=3] (-2,-0.2) -- (-1.4,-0.2) -- (-1,-0.6) -- (-0.7,-0.6);
         \draw[line width=1.5,rounded corners=3] (-0.5,-0.6) -- (-0.2,-0.6) -- (-0.2,0);
         \draw[line width=1.5,rounded corners=3] (2,-0.2) -- (1.4,-0.2) -- (1.2,-0.6) -- (0.2,-0.6) -- (0.2,0);
         \draw[line width=1.5,rounded corners=3] (-0.2,-2) -- (-0.2,-1.4) -- (-0.6,-1) -- (-0.6,0);
         \draw[line width=1.5,rounded corners=3] (0.6,0) -- (0.6,0.6) -- (-0.2,0.6) -- (-0.2,2);
         \foreach \x in {-0.6,-0.2,...,1}
           \fill[black] (\x,0) circle (0.075);
           \foreach \x in {-0.6,-0.2,...,1}
           \fill[black] (\x,-2) circle (0.075);
           \foreach \x in {-0.6,-0.2,...,1}
           \fill[black] (\x,2) circle (0.075);
           \foreach \x in {-0.6,-0.2,...,1}
           \fill[black] (-2,\x) circle (0.075);
           \foreach \x in {-0.6,-0.2,...,1}
           \fill[black] (2,\x) circle (0.075);
     \end{tikzpicture} 
\end{array}
\ee
Here orange lines indicate a Bell-type measurement one the logical qubit (left) and on the central qubit of the 5-qubit chip. The result is the 4-qubit state $|\Psi\rangle$ on the right, with the white space being a box connecting the points in the way prescribed by the logical qubit. Note that the Bell measurement can generate a unitary (orange box) that needs to be undone at the end of the recovery. See~\cite{Melnikov:2022vij} for more details on the teleportation protocol in the topological presentation.

The coefficients of the state $|\Psi\rangle$ in the computational basis can be easily calculated using the pseudounitary representation described in section~\ref{sec:punitary}. For example, the above diagram can be cast in the form of the following ``braid'':
\be
|\Psi\rangle \ = \ \alpha ~ \begin{array}{c}
   \begin{tikzpicture}   
       \newcommand{\y}{0}
       \renewcommand{\y}{4.2}
       \newcommand{\z}{-1.6}
       \fill[olive,opacity=0.5,rounded corners=5] (0,0.05) -- (-0.2,0.05) -- (-0.2,1.55) -- (0.2,1.55) -- (0.2,0.05) -- (0,0.05);
       \fill[olive,opacity=0.5,rounded corners=5] (0,0.05) -- (-0.2+\y,0.05) -- (-0.2+\y,1.55) -- (0.2+\y,1.55) -- (0.2+\y,0.05) -- (0+\y,0.05);
       \fill[olive,opacity=0.5,rounded corners=5] (0,0.05+\z) -- (-0.2,0.05+\z) -- (-0.2,1.55+\z) -- (0.2,1.55+\z) -- (0.2,0.05+\z) -- (0,0.05+\z);
       \fill[olive,opacity=0.5,rounded corners=5] (0+\y,0.05+\z) -- (-0.2+\y,0.05+\z) -- (-0.2+\y,1.55+\z) -- (0.2+\y,1.55+\z) -- (0.2+\y,0.05+\z) -- (0+\y,0.05+\z);
       \draw[line width=1.5] (0,1.4) -- (\y,1.4);
       \draw[line width=1.5] (0,-1.4) -- (\y,-1.4);
       \draw[line width=1.5, rounded corners=3] (0,0.2) -- (0.5,0.2) -- (0.5,-0.2) -- (0,-0.2);
       \draw[line width=1.5, rounded corners=3] (\y,0.2) -- (\y-0.5,0.2) -- (\y-0.5,-0.2) -- (\y,-0.2);
       \draw[line width=1.5, rounded corners=3] (0,-0.6) -- (1,-0.6) -- (1.4,-0.2) -- (1.7,-0.2) -- (1.7,-0.6) -- (1.4,-0.6) -- (1.3,-0.5);
       \draw[line width=1.5, rounded corners=3] (1.1,-0.3) -- (1.0,-0.2) -- (0.7,-0.2) -- (0.7,0.2) -- (1.1,0.2) -- (1.1,0.6) -- (0,0.6);
       \draw[line width=1.5, rounded corners=3] (0,-1.0) -- (2.2,-1.0) -- (2.3,-0.9);
       \draw[line width=1.5, rounded corners=3] (2.5,-0.7) -- (2.6,-0.6) -- (2.8,-0.6) -- (2.9,-0.5);
       \draw[line width=1.5, rounded corners=3] (3.1,-0.3) -- (3.2,-0.2) -- (3.5,-0.2) -- (3.5,0.2) -- (3.1,0.2) -- (3.1,0.6) -- (\y,0.6);
       \draw[line width=1.5, rounded corners=3] (0,1.0) -- (1.6,1.0) -- (1.7,0.9);
       \draw[line width=1.5, rounded corners=3] (1.9,0.7) -- (2.0,0.6) -- (2.9,0.6) -- (2.9,0.2) -- (2.5,0.2) -- (2.5,-0.2) -- (2.8,-0.2) -- (3.2,-0.6) -- (\y,-0.6);
       \draw[line width=1.5, rounded corners=3] (\y,1.0) -- (2.0,1.0) -- (1.6,0.6) -- (1.3,0.6) -- (1.3,0.2) -- (2.3,0.2) -- (2.3,-0.2) -- (1.9,-0.2) -- (1.9,-0.6) -- (2.2,-0.6) -- (2.6,-1) -- (\y,-1);
       \foreach \x in {-1.4,-1.0,...,1.4}
           \fill[black] (0,\x) circle (0.075);
        \foreach \x in {-1.4,-1.0,...,1.4}
           \fill[black] (\y,\x) circle (0.075);
        \draw[orange,line width=1.5, rounded corners=3] (1.5,-0.2) -- (1.7,-0.2) -- (1.7,-0.6) -- (1.5,-0.6);
        \draw[orange,line width=1.5, rounded corners=3] (2.1,-0.2) -- (1.9,-0.2) -- (1.9,-0.6) -- (2.1,-0.6);
        \fill[black] (1.5,-0.2) circle (0.075);
        \fill[black] (1.5,-0.6) circle (0.075);
        \fill[black] (2.1,-0.2) circle (0.075);
        \fill[black] (2.1,-0.6) circle (0.075);
   \end{tikzpicture}
\end{array}
\ + \ \beta ~
\begin{array}{c}
   \begin{tikzpicture}
       \newcommand{\y}{0}
       \renewcommand{\y}{4.2}
       \newcommand{\z}{-1.6}
       \fill[olive,opacity=0.5,rounded corners=5] (0,0.05) -- (-0.2,0.05) -- (-0.2,1.55) -- (0.2,1.55) -- (0.2,0.05) -- (0,0.05);
       \fill[olive,opacity=0.5,rounded corners=5] (0,0.05) -- (-0.2+\y,0.05) -- (-0.2+\y,1.55) -- (0.2+\y,1.55) -- (0.2+\y,0.05) -- (0+\y,0.05);
       \fill[olive,opacity=0.5,rounded corners=5] (0,0.05+\z) -- (-0.2,0.05+\z) -- (-0.2,1.55+\z) -- (0.2,1.55+\z) -- (0.2,0.05+\z) -- (0,0.05+\z);
       \fill[olive,opacity=0.5,rounded corners=5] (0+\y,0.05+\z) -- (-0.2+\y,0.05+\z) -- (-0.2+\y,1.55+\z) -- (0.2+\y,1.55+\z) -- (0.2+\y,0.05+\z) -- (0+\y,0.05+\z);
       \draw[line width=1.5] (0,1.4) -- (\y,1.4);
       \draw[line width=1.5] (0,-1.4) -- (\y,-1.4);
       \draw[line width=1.5, rounded corners=3] (0,0.2) -- (0.5,0.2) -- (0.5,-0.2) -- (0,-0.2);
       \draw[line width=1.5, rounded corners=3] (\y,0.2) -- (\y-0.5,0.2) -- (\y-0.5,-0.2) -- (\y,-0.2);
       \draw[line width=1.5, rounded corners=3] (0,-0.6) -- (1,-0.6) -- (1.4,-0.2) -- (2.3,-0.2) -- (2.3,0.2) -- (1.3,0.2) -- (1.3,0.6) -- (1.6,0.6) -- (2.0,1.0) -- (\y,1.0);
       \draw[line width=1.5, rounded corners=3] (1.1,-0.3) -- (1.0,-0.2) -- (0.7,-0.2) -- (0.7,0.2) -- (1.1,0.2) -- (1.1,0.6) -- (0,0.6);
       \draw[line width=1.5, rounded corners=3] (0,-1.0) -- (2.2,-1.0) -- (2.3,-0.9);
       \draw[line width=1.5, rounded corners=3] (2.5,-0.7) -- (2.6,-0.6) -- (2.8,-0.6) -- (2.9,-0.5);
       \draw[line width=1.5, rounded corners=3] (3.1,-0.3) -- (3.2,-0.2) -- (3.5,-0.2) -- (3.5,0.2) -- (3.1,0.2) -- (3.1,0.6) -- (\y,0.6);
       \draw[line width=1.5, rounded corners=3] (0,1.0) -- (1.6,1.0) -- (1.7,0.9);
       \draw[line width=1.5, rounded corners=3] (1.9,0.7) -- (2.0,0.6) -- (2.9,0.6) -- (2.9,0.2) -- (2.5,0.2) -- (2.5,-0.2) -- (2.8,-0.2) -- (3.2,-0.6) -- (\y,-0.6);
       \draw[line width=1.5, rounded corners=3] (1.3,-0.5) -- (1.4,-0.6) -- (2.2,-0.6) -- (2.6,-1) -- (\y,-1);
       \foreach \x in {-1.4,-1.0,...,1.4}
           \fill[black] (0,\x) circle (0.075);
        \foreach \x in {-1.4,-1.0,...,1.4}
           \fill[black] (\y,\x) circle (0.075);
           \draw[orange,line width=1.5, rounded corners=3] (1.5,-0.2) -- (2.1,-0.2);
        \draw[orange,line width=1.5, rounded corners=3] (2.1,-0.6) -- (1.5,-0.6);
        \fill[black] (1.5,-0.2) circle (0.075);
        \fill[black] (1.5,-0.6) circle (0.075);
        \fill[black] (2.1,-0.2) circle (0.075);
        \fill[black] (2.1,-0.6) circle (0.075);
   \end{tikzpicture}
\end{array}\,.
\ee
The state is a linear combination of two diagrams with two different routings (shown in orange) corresponding to basis~(\ref{basis}). In order for the above to be a state in the tensor product of four qubits, the highlighted groups of points should be projected on the orthogonal basis~(\ref{ONbasis}).

Using the matrix representation of section~\ref{sec:punitary}, one finds the following table of coefficients
\be
\label{coeffs}
\begin{tabular}{||c|c|c|c||}
  \hline  $a_{0000}$  & $a_{0001}$ & $a_{0010}$ & $a_{0011}$ \\
  \hline  $\ds \frac{\alpha d+\beta}{d^3}$ & $\ds \frac{\sqrt{\Delta}(\alpha d+\beta)}{d^3}$ & $\ds\frac{A^8\sqrt{\Delta}(\alpha d+\beta)}{d^3}$ & $\ds-\frac{A^8(\alpha d+\beta)}{d^3}$\\ \hline 
  \hline  $a_{0100}$  & $a_{0101}$ & $a_{0110}$ & $a_{0111}$ \\
  \hline  $\ds \frac{\sqrt{\Delta}\beta}{A^{8}d^3}$ & $\ds -\frac{\beta}{A^{8}d^3}$ & $\ds  \frac{{\Delta}\beta}{d^3}$ & $\ds \frac{\beta}{\sqrt{\Delta}d^3}$\\ \hline
  \hline  $a_{1000}$  & $a_{1001}$ & $a_{1010}$ & $a_{1011}$ \\
  \hline  $\ds  \frac{\sqrt{\Delta}\beta}{d^3} $ & $\ds \frac{{\Delta}\beta}{d^3}$ & $\ds -\frac{A^8\beta}{d^3} $ & $\ds \frac{A^{8}\beta}{\sqrt{\Delta}d^3}$\\ \hline
  \hline  $a_{1100}$  & $a_{1101}$ & $a_{1110}$ & $a_{1111}$ \\
  \hline  $\ds \frac{\alpha d -A^{-8}\beta}{d^3}$ & $\ds -\frac{\alpha d -A^{-8}\beta}{\sqrt{\Delta} d^3}$ & $\ds -\frac{A^8(\alpha d -A^{-8}\beta)}{\sqrt{\Delta} d^3} $ & $\ds -\frac{A^2\alpha  -\beta}{d} - \frac{A^4\alpha d +\beta}{\Delta d^3}$\\ \hline
\end{tabular}
\ee
where $\Delta = d^2-1$.

\end{enumerate}

\end{appendix}


\end{document}